\documentclass[a4paper, 12pt]{article}

\pdfoutput = 1

\usepackage{styleBuding}
\usepackage{bm}
\usepackage{color,listings}
\usepackage[usenames,dvipsnames,svgnames,table]{xcolor}
\usepackage{amsthm, amsmath}
\usepackage{amssymb}
\usepackage{mathrsfs}
\usepackage{graphicx}
\usepackage{slashed}
\usepackage{soul}
\usepackage{multirow}
\usepackage{subfigure}
\usepackage{diagbox}
\usepackage{siunitx} 
\usepackage{url} 
\usepackage[utf8]{inputenc}
\usepackage[figuresright]{rotating}
\definecolor{back}{HTML}{F8F8F8}
\usepackage{epstopdf}
\usepackage{lipsum}
\usepackage{appendix}
\usepackage{booktabs}    

\usepackage{array}

\newcommand{\rom}[1]{\uppercase\expandafter{\romannumeral #1\relax}}

\allowdisplaybreaks
\DeclareUnicodeCharacter{2212}{-}

\title{Viability of Sub-TeV Higgsino Dark Matter with Nearly Mass-Degenerate Sleptons}
\author{Yuanfang Yue$^a$, Yuetao Wang$^{a}$}
\affiliation{ $^a$ School of Physics, Henan Normal University, Xinxiang 453007, China}
\emailAdd{yueyuanfang@htu.edu.cn}
\emailAdd{wangyuetao@stu.htu.edu.cn}

\abstract{The higgsino-like neutralino is a compelling dark matter candidate motivated by both cosmology and naturalness considerations. While a pure higgsino typically requires a mass of around $1.1~\mathrm{TeV}$ to satisfy the observed thermal relic abundance, the presence of light sleptons can significantly alter this requirement. In this work, we revisit higgsino dark matter within the Minimal Supersymmetric Standard Model (MSSM), focusing on scenarios with slepton coannihilation. We find that the presence of nearly mass-degenerate sleptons in the thermal bath can allow the higgsino mass to be as light as $\sim 400$ GeV while satisfying relic density constraints. We explicitly contrast the impact of recent direct detection updates: the LZ-2022 limits raise this lower bound to approximately $450~\mathrm{GeV}$, while the stringent LZ-2024 constraints further shift the viable mass floor to $\sim 500~\mathrm{GeV}$. Crucially, we demonstrate that the direct detection sensitivity is strongly dependent on the relative signs of the gaugino mass parameters $M_1$ and $M_2$. We find that scenarios with $M_1, M_2 > 0$ are fully excluded by LZ-2024. Conversely, configurations with opposite signs ($M_1/M_2 < 0$) remain broadly viable, as destructive interference in the neutralino-Higgs coupling efficiently suppresses the spin-independent cross section. Finally, we delineate the remaining viable parameter space for both the opposite-sign cases and the specific configurations with negative $M_1$ and $M_2$.}

\begin{document}
    \maketitle
    \flushbottom

    \section{Introduction}
    \label{section:intro}
    
    For decades, compelling evidence for dark matter (DM) has accumulated from a wide range of astrophysical and cosmological observations~\cite{zwicky1942large,zwicky1959clusters,zwicky1960spectra,trimble1987existence,sanders2010dark,Planck:2018vyg,pecontal2009review}, yet its fundamental nature remains one of the most profound puzzles in modern physics. Weakly Interacting Massive Particles (WIMPs) arising from new physics at the TeV scale provide an elegant explanation, and among the most well-motivated frameworks is Supersymmetry (SUSY). In the Minimal Supersymmetric Standard Model (MSSM)~\cite{Nilles:1983ge,Haber:1984rc,Gunion:1984yn,csaki1996minimal,Martin:1997ns,haber1998status}, the lightest neutralino, a mixture of the bino, wino, and higgsinos, serves as a natural DM candidate.
    
    However, the pure states of neutralino DM face opposing phenomenological challenges. A bino-like neutralino typically possesses a small annihilation cross-section, leading to an over-abundance of relic density unless specific mechanisms, such as coannihilation with scalars or resonant annihilation, enhance the annihilation rate~\cite{Edsjo:1997bg}. Conversely, wino-like and pure higgsino-like neutralinos annihilate efficiently via electroweak interactions. A wino-like DM is typically under-abundant unless its mass is around $2.5~\text{TeV}$~\cite{Beneke:2016jpw}, and it is severely constrained by indirect and direct detection searches~\cite{Cohen:2013ama,Fan:2013faa,Safdi:2025sfs,Aghaie:2025iyn}. Similarly, a pure higgsino is typically under-abundant unless its mass is raised to approximately $1.1~\text{TeV}$~\cite{Delgado:2020url,Kowalska:2018toh,Chattopadhyay:2005mv}. This heavy thermal mass target renders the pure higgsino largely inaccessible to current collider experiments.
    
    Despite these challenges, the higgsino remains a prime DM candidate in both particle physics and cosmology communities~\cite{1995Supersymmetric,Kane:1996dd,dreesLightHiggsinoDark1997,Masip:2005fv,Wang:2005kf,Hall:2009aj,Sinha:2012kq,Mayes:2014wda,Kowalska:2018toh,Delgado:2020url,Wang:2024ozr,Evans:2022gom,Nagata:2014wma,Aparicio:2016qqb}. Theoretically, it is appealing due to its connection to electroweak naturalness via the $\mu$ parameter~\cite{Baer:2012uy,Ellwanger:2018zxt,Baer:2018rhs,Bae:2017hlp,baerRadiativeNaturalSupersymmetry2013a,Han:2019vxi} and its role as a realization of minimal dark matter~\cite{Cirelli:2005uq,Bottaro:2022one,Baumgart:2025dov}, which can easily be embedded in frameworks like Split-SUSY~\cite{Fox:2014moa,Arkani-Hamed:2004ymt,Giudice:2004tc,Arkani-Hamed:2004zhs}. The key difficulty remains its large annihilation cross-section in the sub-TeV regime, which results in a relic density far below the observed value~\cite{Arkani-Hamed:2006wnf,Delgado:2020url,Chattopadhyay:2005mv,Chakraborti:2014fha,Fox:2014moa,Shafi:2023sqa,Mummidi:2018myd}.
    
    In this work, we focus on a realistic MSSM scenario where the dark matter is higgsino-dominated but possesses non-negligible bino and wino admixtures. We investigate how the synergy between this three-component mixing and slepton coannihilation~\cite{Griest:1990kh,Edsjo:1997bg,Baker:2015qna} enables the correct relic density to be achieved for DM masses significantly below the 1 TeV thermal target—specifically extending down to $\sim 400~\text{GeV}$. Reviving this sub-TeV window is crucial as it brings the higgsino candidate back within the reach of experimental searches. It is important to distinguish this setup from the standard well-tempered neutralino scenario~\cite{Arkani-Hamed:2006wnf}. While well-tempered models rely solely on fine-tuning the mixing angles to satisfy relic density, our approach utilizes slepton coannihilation as an essential mechanism. This opens up a broader and lighter parameter space that would otherwise be forbidden in a purely well-tempered framework.
    
    Higgsino coannihilation with sleptons has been previously investigated, for instance in Ref.~\cite{Chakraborti:2017dpu}, which found a lower mass limit of around $600~\text{GeV}$ under LUX 2016 constraints with fixed $\tan\beta \approx 10$. In this work, we revisit this scenario by performing a comprehensive scan over the parameter space, allowing $\tan\beta$ to vary and applying the latest, far more restrictive limits from the LZ 2022 and LZ 2024 experiments~\cite{LZ:2022lsv,LZ:2024zvo}. We find that the DM mass can still be as low as $\sim 500~\text{GeV}$, provided that the model resides in specific interference regions determined by the relative signs of $M_1$ and $M_2$.
    
    The presence of bino and wino components is particularly consequential for these direct detection limits. The spin-independent scattering rate is governed by the interference between the bino and wino contributions to the neutralino-Higgs coupling. As hinted above, the relative signs of the mass parameters $M_1$ and $M_2$ dictate whether this interference is constructive or destructive. This suppression mechanism is analogous to the well-known blind spot phenomenon discussed in well-tempered scenarios~\cite{Cheung:2012qy,Han:2016qtc,Huang:2014xua}. However, a key distinction in our work is that the relic density is satisfied via slepton coannihilation rather than pure gaugino-higgsino mixing. Consequently, we identify viable ``blind spot-like'' regions in parameter space that are distinct from those found in standard well-tempered models, allowing the dark matter to evade even the stringent constraints from the LZ-2024 experiment.
    
    It is worth noting that in the extreme decoupling limit, where gaugino masses are very heavy, the mass splittings between higgsino states can become extremely small, opening up the possibility of inelastic scattering signatures~\cite{Feldstein:2010su,Tucker-Smith:2001myb}, which are themselves tightly constrained~\cite{Graham:2024syw}. However, in the mixed scenario we consider, our focus remains on the elastic scattering interference effects which are most relevant for current direct detection limits.
    
    This work is organized as follows. In Section~\ref{theory-section}, we review the theoretical framework and establish our notation. In Section~\ref{numerical study}, we detail our numerical scan strategy and present the main results. We showcase several representative benchmark points in Section~\ref{BMP}. Finally, we summarize our findings and conclude in Section~\ref{conclusion}.

\section{\label{theory-section}Theoretical Preliminaries} 

Dark matter physics within the MSSM is inextricably linked to the Higgs, neutralino, and chargino sectors. In this section, we briefly review the relevant theoretical framework, outlining the mass spectra and mixing structures of these sectors, which are fundamental to the calculation of the dark matter relic density and direct detection cross sections.

\subsection{\label{Section-Model} The Higgs and Neutralino Sectors}

The Higgs sector of the MSSM consists of two Higgs doublets $H_u$ and $H_d$. After electroweak symmetry breaking (EWSB), the neutral components of these doublets acquire vacuum expectation values (VEVs), denoted as $v_u$ and $v_d$ respectively. Phenomenological studies conventionally define the ratio of these VEVs as $\tan \beta = v_u / v_d$. To obtain the mass eigenstates, the neutral scalar fields are expanded around their VEVs as $H_{k}^0 = v_{k} + \frac{1}{\sqrt{2}}(S_{k} + i P_{k})$ for $k=d,u$. The CP-even Higgs mass eigenstates $h$ and $H$ are obtained by rotating the interaction eigenstates $S_d$ and $S_u$ by the mixing angle $\alpha$

\begin{equation}
\begin{pmatrix} h \\ H \end{pmatrix} = \begin{pmatrix} -\sin\alpha & \cos\alpha \\ \cos\alpha & \sin\alpha \end{pmatrix} \begin{pmatrix} S_d \\ S_u \end{pmatrix}
\label{eq:higgs_mixing}
\end{equation}

Here, $h$ is identified as the SM-like Higgs boson with a mass of approximately 125 GeV, while $H$ represents the heavy CP-even Higgs boson. In this work, we assume a decoupling limit where the heavier Higgs bosons are sufficiently massive to be removed from the low-energy spectrum, thereby focusing our analysis on the properties and interactions of the SM-like Higgs boson.

The neutralino sector is composed of four Majorana fermions resulting from the mixing of the neutral bino, wino, and higgsino fields. In the gauge eigenstate basis $\psi_j^0 = (\tilde{B}, \tilde{W}^0, \tilde{H}_d^0, \tilde{H}_u^0)$, the neutralino mass matrix is given by \cite{Martin:1997ns}:

\begin{equation}
M_{\tilde{\chi}^0} =
\begin{pmatrix}
M_1 & 0 & -\cos\beta \sin\theta_W m_Z & \sin\beta \sin\theta_W m_Z \\
0 & M_2 & \cos\beta \cos\theta_W m_Z & -\sin\beta \cos\theta_W m_Z \\
-\cos\beta \sin\theta_W m_Z & \cos\beta \cos\theta_W m_Z & 0 & -\mu \\
\sin\beta \sin\theta_W m_Z & -\sin\beta \cos\theta_W m_Z & -\mu & 0
\end{pmatrix}
\label{eq:neutralino_mass_matrix}
\end{equation}
Here, $M_1$ and $M_2$ are the soft SUSY-breaking gaugino mass parameters associated with the $U(1)_Y$ and $SU(2)_L$ gauge groups respectively; $\mu$ is the higgsino mass parameter; $\theta_W$ is the weak mixing angle; and $m_Z$ is the $Z$ boson mass.

The neutralino mass matrix $M_{\tilde{\chi}^0}$ is diagonalized by a unitary matrix $N$ to yield the physical masses $m_{\tilde{\chi}_i^0}$

\begin{equation}
N^* M_{\tilde{\chi}^0} N^{-1} = \text{diag}(m_{\tilde{\chi}_1^0}, m_{\tilde{\chi}_2^0}, m_{\tilde{\chi}_3^0}, m_{\tilde{\chi}_4^0}).
\label{eq:neutralino_masses}
\end{equation}
The mass eigenstates $\tilde{\chi}_i^0$ (ordered by mass such that $m_{\tilde{\chi}_1^0} < \dots < m_{\tilde{\chi}_4^0}$) are related to the gauge eigenstates via

\begin{equation}
\tilde{\chi}_i^0 = N_{ij} \psi_j^0.
\label{eq:neutralino_eigenstates}
\end{equation}

In the MSSM, the lightest neutralino $\tilde{\chi}_1^0$ acts as a compelling dark matter candidate. Its composition is expressed as

\begin{equation}
\tilde{\chi}_1^0 = N_{11} \tilde{B} + N_{12} \tilde{W}^0 + N_{13} \tilde{H}_d^0 + N_{14} \tilde{H}_u^0.
\label{eq:lsp_composition}
\end{equation}
The physical nature of the lightest neutralino is determined by its dominant component. It is classified as bino-like, wino-like, or higgsino-like if the largest contribution comes from $|N_{11}|^2$, $|N_{12}|^2$, or $|N_{13}|^2 + |N_{14}|^2$  respectively. This work specifically investigates the higgsino-like scenario.

The chargino sector consists of two Dirac fermions, formed by the mixing of the charged wino and higgsino components. In the basis $\psi^+ = (\tilde{W}^+, \tilde{H}_u^+)$ and $\psi^- = (\tilde{W}^-, \tilde{H}_d^-)$, the chargino mass matrix reads \cite{Martin:1997ns}

\begin{equation}
M_{\tilde{\chi}^\pm} = \begin{pmatrix}
M_2 & \sqrt{2} \sin\beta m_W \\
\sqrt{2} \cos\beta m_W & \mu
\end{pmatrix},
\label{eq:chargino_mass_matrix}
\end{equation}
where $m_W$ is the $W$ boson mass. The chargino mass eigenstates $\tilde{\chi}_i^\pm$ ($i=1,2$) are related to the gauge eigenstates through two unitary matrices $U$ and $V$

\begin{equation}
\tilde{\chi}_i^+ = V_{ij} \psi_j^+ , \quad 
\tilde{\chi}_i^- = U_{ij} \psi_j^-. \label{eq:chargino_eigenstates_minus}
\end{equation}
The diagonalization condition is given by:

\begin{equation}
U^* M_{\tilde{\chi}^\pm} V^{-1} = \text{diag}(m_{\tilde{\chi}_1^\pm}, m_{\tilde{\chi}_2^\pm}).
\label{eq:chargino_masses}
\end{equation}

In the limit $|\mu| \ll M_1, M_2$, the lightest neutralino becomes predominantly higgsino-like.Consequently, the masses of $\tilde{\chi}_1^0$, $\tilde{\chi}_2^0$, and $\tilde{\chi}_1^\pm$ become nearly degenerate, all approximating the value $|\mu|$. This mass degeneracy is critical for the relic density calculation, as it enhances co-annihilation channels.

However, even in the higgsino-dominated regime, the bino and wino soft-breaking mass parameters $M_1$ and $M_2$ play a significant role in dark matter phenomenology. They induce mixing that affects the couplings to the Higgs and $Z$ bosons, thereby significantly impacting the relic density and direct detection cross sections. The interplay between these gaugino masses and the higgsino parameter $\mu$—especially under the stringent constraints from LZ2024—will be a key focus of our analysis.

\subsection{\label{RD} Dark Matter Relic Density}

The relic abundance of dark matter serves as a pivotal observable for constraining the parameter space of BSM physics. In the standard thermal freeze-out paradigm, the relic density is governed by the annihilation interactions of dark matter particles in the early universe. The time evolution of the dark matter number density $n$ is described by the Boltzmann equation \cite{Griest:1990kh,1995Supersymmetric,Bertone:2004pz}
\begin{equation}
\frac{dn}{dt} + 3Hn = -\langle \sigma v \rangle (n^2 - n_{\mathrm{eq}}^2),
\end{equation}
where $H$ is the Hubble parameter, $\langle \sigma v \rangle$ is the thermally averaged annihilation cross section multiplied by the relative velocity, and $n_{\mathrm{eq}}$ is the equilibrium number density. In the absence of co-annihilation effects, the relic density can be approximated as
\begin{equation}
	\Omega h^2 \simeq 0.12 \left(\frac{80}{g_*}\right)^{1/2} \left(\frac{x_F}{25}\right) \left(\frac{2.3 \times 10^{-26} \, \mathrm{cm}^3/\mathrm{s}}{\langle\sigma v\rangle_{x_F}}\right),
\end{equation}
where $g_*$ is the effective number of relativistic degrees of freedom at freeze-out, and $x_F \equiv m_{\mathrm{DM}}/T_F$, with $T_F$ being the freeze-out temperature \cite{Baum:2017enm}.

However, when the dark matter candidate is part of a compressed mass spectrum, co-annihilation effects become significant and must be incorporated. This occurs when other sparticles $\chi_i$ have masses nearly degenerate with the lightest supersymmetric particle (LSP) $\chi_1$ (where $m_{\chi_1} < m_{\chi_2} < \dots$).\footnote{In our specific context, $\chi_1$ corresponds to the LSP $\tilde{\chi}_1^0$, while $\chi_i$ ($i>1$) represents the NLSPs, such as $\tilde{\chi}_1^\pm$ and $\tilde{\chi}_2^0$, which exhibit a small mass splitting with the LSP.} Under these conditions, the standard thermally averaged cross section in the Boltzmann equation is replaced by an effective annihilation cross section $\langle \sigma v \rangle_{\mathrm{eff}}$ defined as \cite{Griest:1990kh}

\begin{equation}
	\langle \sigma v \rangle_{\mathrm{eff}}
	= \sum_{i,j} \langle \sigma_{ij} v \rangle \, \frac{g_i g_j}{g_{\mathrm{eff}}^{2}}
	   (1+\Delta_i)^{3/2}(1+\Delta_j)^{3/2}
	   \times \exp\!\left[-x\left(\Delta_i+\Delta_j\right)\right],
\end{equation}
where the summation runs over all relevant co-annihilating species. Here $\langle \sigma_{ij} v \rangle$ denotes the thermally averaged cross section for the process $\chi_i \chi_j \to X X^\prime$ (with $X, X^\prime$ being SM particles). The mass splitting parameter is defined as $\Delta_i = (m_{\chi_i} - m_{\chi_1})/m_{\chi_1}$, and $g_i$ represents the internal degrees of freedom of particle $\chi_i$. The effective degrees of freedom $g_{\mathrm{eff}}$ is given by \cite{Griest:1990kh}
\begin{equation}
	g_{\mathrm{eff}} = \sum_i g_i (1 + \Delta_i)^{3/2} \exp(-x \Delta_i).
	\label{geff}
\end{equation}
Equation~(\ref{geff}) implies that contributions from co-annihilating partners are exponentially suppressed by their mass splitting $\Delta_i$. Therefore, only states that are nearly degenerate with the LSP play a significant role in determining the effective cross section and the resulting relic density.

A distinctive feature of our analysis is the inclusion of light sleptons in the spectrum. In standard scenarios where co-annihilation is restricted solely to the $\tilde{\chi}_2^0$ and $\tilde{\chi}_1^\pm$, it is rather difficult to realize a relatively light Higgsino dark matter candidate consistent with the observed relic density. However, in this work, we explicitly explore the parameter space where sleptons are sufficiently light to participate in the freeze-out process. The introduction of these additional slepton co-annihilation channels significantly modifies the effective annihilation cross section. A key objective of this study is to investigate how these channels impact the lower bound of the dark matter mass, specifically determining whether the presence of light sleptons renders lighter Higgsino dark matter phenomenologically viable under the stringent constraints of LZ2024.

\subsection{\label{DMDD} Dark Matter Direct Detection}

Direct detection experiments aim to observe dark matter particles through their elastic scattering off atomic nuclei in terrestrial detectors. This scattering process generally comprises two distinct components: spin-independent (SI) and spin-dependent (SD) interactions.

The dominant contribution to the SI scattering cross section typically arises from the $t$-channel exchange of CP-even Higgs bosons, assuming that squarks are sufficiently heavy. The SI cross section for dark matter scattering off a nucleon $N$ can be expressed as \cite{Bertone:2004pz}

\begin{equation}
\sigma^{\mathrm{SI}} = \frac{4}{\pi} \left(\frac{m_{\tilde{\chi}_1^0}  m_N}{m_{\tilde{\chi}_1^0} + m_N}\right)^2 f_N^2,
\end{equation}
where $m_{\tilde{\chi}_1^0}$ is the dark matter mass, $m_N$ is the nucleon mass, and $f_N$ represents the effective scalar coupling between the dark matter particle and the nucleon. This effective coupling is given by

\begin{equation}
	f_N = \sum_{i} \frac{C_{\tilde{\chi}_1^0 \tilde{\chi}_1^0 h_i} \, C_{h_i N N}}{m_{h_i}^2},
\end{equation}
where the summation runs over all CP-even Higgs eigenstates. In our framework, assuming the heavier Higgs bosons are decoupled, only the SM-like Higgs boson $h$ contributes significantly.   In the higgsino-dominated limit, the tree-level coupling of the lightest neutralino to the SM-like Higgs is approximated as~\cite{Hisano:2004pv,Martin:2024ytt}
\begin{equation}
  C_{\tilde{\chi}_{1}^{0}\tilde{\chi}_{1}^{0}h} \approx \mp \frac{1}{2}m_W
  \left[\frac{\tan^2 \theta_W}{M_1-|\mu|}+\frac{1}{M_2-|\mu|}\right](1\pm\sin2\beta)\,,\label{eq:DM_HiggsCoupling}
\end{equation}
with the upper (lower) sign corresponding to $\mu>0$ ($\mu<0$). For $|M_{1,2}|\gg|\mu|$, the relative signs of $M_1$ and $M_2$ control the interference in the bracket, while the coupling decouples for heavy gauginos.

The SD scattering process is primarily mediated by the $t$-channel exchange of the $Z$ boson in the heavy-squark limit. To facilitate a direct comparison with the SI case, we use the same denominator structure and write the relevant axial-vector coupling as~\cite{Hisano:2004pv,Martin:2024ytt}
\begin{equation}
  C_{\tilde{\chi}_{1}^{0}\tilde{\chi}_{1}^{0}Z} \approx \mp \frac{1}{2\mu}m_W^2
  \left(\frac{\tan^2 \theta_W}{M_1-|\mu|}+\frac{1}{M_2-|\mu|}\right)\cos2\beta\,.\label{eq:DM_ZCoupling}
\end{equation}
This coupling shows a parametric dependence analogous to the SI case: it is suppressed by heavy gauginos and is enhanced (suppressed) for $M_1$ and $M_2$ with the same (opposite) signs.

In summary, the direct detection sensitivity is critically dependent on the interplay between $M_1$ and $M_2$. Scenarios with opposite-sign gaugino masses benefit from suppressed effective couplings, offering a mechanism to evade detection. Conversely, the constructive interference in same-sign scenarios makes them highly susceptible to experimental bounds. Consequently, the stringent constraints from LZ2024 are expected to exclude a significant portion of the same-sign parameter space, a feature we will examine closely in the following numerical analysis.

\section{\label{numerical study}Numerical Study}

In this section, we present our numerical study of Higgsino-like dark matter within the framework of the MSSM. We begin by outlining our research strategy, detailing the parameter space scan and the applied constraints. Subsequently, we present and discuss our numerical results, emphasizing the impact of coannihilation effects on the lower bound of DM mass and the roles of $M_1$ and $M_2$ in determining the direct detection cross-sections.

\subsection{\label{scan}Research Strategy}

We explore the parameter space using the \texttt{EasyScan\_HEP} package \cite{Shang:2023gfy}, which implements the parallelized \texttt{MultiNest} algorithm \cite{feroz2009multinest} for efficient sampling. The relevant model parameters are varied within the ranges specified in Table \ref{ScanRange}. The rationale behind these parameter choices is detailed below

\begin{itemize}
  \item {\bf Higgsino Mass Parameter ($\mu$)} The Higgsino mass parameter $\mu$ is scanned from 400 GeV to 1200 GeV. This interval covers the spectrum from the canonical thermal relic mass of $\sim 1.1$ TeV down to the sub-TeV regime, allowing us to investigate the viability of lighter Higgsino dark matter. To efficiently probe the phenomenologically interesting lower mass region while satisfying the stringent relic density and direct detection constraints, we adopt a segmented scanning strategy. The sub-TeV range is divided into multiple intervals: $\mu \in (400,500), (500,600), \ldots, (900,1000)$ GeV, followed by a final segment of $(1000,1200)$ GeV. This approach ensures sufficient sampling density in the low-mass region where co-annihilation effects are most critical.

  \item {\bf $\tan \beta$ and Trilinear Parameters ($A_t, A_b$)} The ratio of vacuum expectation values $\tan \beta$ is varied between 3 and 30. The lower bound is chosen to facilitate the realization of the 125 GeV Higgs mass and to satisfy LEP constraints on the Higgs sector \cite{Mahmoudi:2010xp}. The upper bound is restricted to avoid color-breaking minima and to ensure vacuum stability \cite{King:2012is,Arbey:2012ax}. The trilinear soft-breaking parameter $A_t$ is scanned in the range of 2500 GeV to 5000 GeV to provide sufficient radiative corrections for the SM-like Higgs mass. For simplicity, we assume the relation $A_b = A_t$.

  \item {\bf CP-odd Higgs Mass ($m_A$)} The mass parameter $m_A$ is constrained to be greater than 500~GeV to evade exclusion limits from heavy Higgs searches \cite{ATLAS:2019nkf}. The upper bound of 2000~GeV on $m_A$ is a scan-range choice adopted for computational efficiency and to focus on the phenomenologically relevant non-decoupling regime \cite{Arbey:2012ax,Mahmoudi:2018xml}. Note that in our scan procedure, $m_A$ serves as a tree-level input parameter defined at the scale $Q = 1\text{ TeV}$. We use it to calculate the soft-breaking term $B_{\mu}(Q)$ via the tree-level relation $m_A^2 = 2B_{\mu} / \sin(2\beta)$\cite{Martin:1997ns}, which is then passed to the SPheno executable as the actual running input. Consequently, the physical mass of the CP-odd Higgs boson presented in our results is the full loop-corrected pole mass (e,g. entry $A$ in Table \ref{BMP12} and \ref{BMP34}), which may deviate from the tree-level input due to radiative corrections.

  \item {\bf Gaugino Mass Parameters ($M_1, M_2$)} The soft-breaking gaugino masses $M_1$ and $M_2$ are varied freely within the range of $[-6000, 6000]$ GeV. This wide range allows for a comprehensive exploration of their impact on the dark matter relic density and direct detection cross sections, particularly through their influence on the neutralino mixing and couplings.

  \item {\bf Slepton Mass Splitting ($\Delta_L$)} To specifically investigate slepton co-annihilation effects, we parametrize the slepton spectrum relative to the LSP mass. We assume a flavor-diagonal slepton sector with no intergenerational mixing. We define a dimensionless mass splitting parameter $\Delta_L$, such that the soft-breaking mass squared parameters for the left-handed sleptons, right-handed sleptons, and sneutrinos are defined as
  \begin{eqnarray}
      (m_{\tilde{L}}^2)_{ij}(Q) = (m_{\tilde{e}}^2)_{ij}(Q) = \left[ |\mu(Q)| (1 + \Delta_L) \right]^2 \, \delta_{ij} \,,
  \end{eqnarray}
  where $\delta_{ij}$ is the Kronecker delta, meaning all off-diagonal entries vanish. All inputs are $\overline{\text{DR}}$ running parameters defined at the scale $Q = 1 ~\text{TeV}$. $\Delta_L$ is varied from 0 to 0.35, as co-annihilation becomes negligible for larger mass splittings. Other soft-breaking masses, including squark masses and the gluino mass parameter $M_3$, are decoupled by fixing them at 3 TeV to isolate the physics of the electroweakino-slepton sector.
\end{itemize}

The parameter ranges defined above represent a targeted estimation based on physical considerations, designed to maximize the likelihood of finding viable points that satisfy all experimental constraints. With these ranges established, we construct the global likelihood function as follows

\begin{equation}
\mathcal{L} \equiv \mathcal{L}_{\Omega h^2} \times \mathcal{L}_{\mathrm{LZ2022}} \times \mathcal{L}_{\mathrm{const}}.
\end{equation}

This total likelihood function is a product of individual likelihoods corresponding to the dark matter relic density, direct detection limits, and other physical constraints.

For the relic density likelihood, $\mathcal{L}_{\Omega h^2}$, we adopt a Gaussian distribution centered at the Planck observed value $\Omega h^2 = 0.12$, with a conservative theoretical uncertainty of 10\% \cite{Aghanim:2018eyx}.
The direct detection likelihood $\mathcal{L}_{\mathrm{LZ2022}}$ is modeled as a one-sided Gaussian with a mean of zero and a variance defined by $\delta_{\sigma} = \sigma_{\mathrm{UL}}^2/1.64^2 + (0.2 \sigma)^2$ (incorporating a 20\% theoretical uncertainty), where $\sigma_{\mathrm{UL}}$ denotes the 90\% confidence level upper limit from the LZ 2022 experiment \cite{LZ:2022lsv}.

The term $\mathcal{L}_{\mathrm{const}}$ enforces hard constraints by assigning a negligible likelihood to parameter points that violate any of the following conditions

\begin{itemize}
  \item {\bf Higgs Sector Constraints} The SM-like Higgs boson mass is required to fall within the window of $125 \pm 3$ GeV (i.e., 122--128 GeV), accounting for both experimental measurements and theoretical uncertainties in the spectrum calculation. Furthermore, constraints from Higgs searches and signal strength measurements are applied using HiggsBounds-5.11.3 \cite{2010hb,2011hb,2014hb}, HiggsSignals-2.6.1\cite{2015applying,2014hs,HS2013xfa,HSConstraining2013hwa,HBHS2012lvg}, and HiggsTools-2.2.0 \cite{Bahl:2022igd}. We require all accepted points to pass the exclusion checks provided by these tools.

  \item {\bf Flavor Physics Constraints} Flavor observables, including $\mathrm{BR}(B\to X_s \gamma)$ and $\mathrm{BR}(B_s \to \mu^+ \mu^-)$, are calculated using FlavorKit \cite{Porod:2014xia}. We require these predictions to be consistent with experimental values \cite{pdg2020} within a $2\sigma$ confidence interval.

  \item {\bf Higgsino Composition} To ensure the analysis focuses on the higgsino-dominated scenario, we impose a filter requiring the higgsino fraction of the LSP, defined as $|N_{13}|^2 + |N_{14}|^2$, to be greater than 0.5.
\end{itemize}

To facilitate these calculations, we generated model files for SPheno-4.0.5\cite{Porod_2012,Porod:2003um,Porod:2011nf} and MicrOMEGAs-5.0.4 \cite{Barducci:2016pcb,Belanger_2009,Belanger:2001fz,Belanger:2010pz,Belanger:2014hqa,Belanger:2018mqt} using SARAH-4.15.4\cite{Staub:2008uz,Staub:2012pb,Staub:2013tta,Staub:2015kfa}. For each input point, SPheno computes the mass spectrum and mixing matrices, while MicrOMEGAs is utilized to evaluate the dark matter relic density and both SI and SD scattering cross sections.Although our numerical scan relies on these standard tree-level evaluations, we note that electroweak radiative corrections can significantly enhance both the spin-independent and spin-dependent cross-sections for Higgsino-like dark matter \cite{Bisal:2023fgb,Bisal:2024ezn}. Consequently, our applied direct detection constraints can be regarded as conservative, as the inclusion of such loop effects would render the actual LZ2024 limits even more stringent.

\begin{table}
	\centering
	\vspace{0.3cm}
	\resizebox{0.7\textwidth}{!}{
	\begin{tabular}{c|c|c|c|c|c}
	  \hline
	  Parameter & Prior & Range & Parameter & Prior & Range \\
	  \hline
	  $\mu/\mathrm{GeV}$ & Flat & $400$--$1200$ &   $\tan\beta$ & Flat & $3$--$30$ \\
	  $m_A/\mathrm{GeV}$ & Flat & $500$--$2000$ &	             $M_1/\mathrm{GeV}$ & Flat & $-6000$--$6000$\\
	  $A_t/\mathrm{GeV}$ & Flat & $2500$--$5000$ &             $M_2/\mathrm{GeV}$ & Flat & $-6000$--$6000$ \\
	  $\Delta_L/\mathrm{GeV}$ & Flat & $0$--$0.35$ & - & - & -  \\ 
	  \hline
	\end{tabular}}
	\caption{Scan ranges for the input parameters explored in this study. All parameters are defined as $\overline{\text{DR}}$ running inputs at the scale $Q = 1 \text{ TeV}$. For simplicity, several parameters are held fixed: all squark soft-breaking masses are set to 2 TeV, the gluino mass parameter $M_3$ is 3 TeV, and we assume $A_t = A_b$. The diagonal charged-lepton trilinear parameters are set to zero ($A_e(Q) = 0$) by default in our main scan, though the impact of non-zero $A_e$ is further examined in Appendix A (Scenario C). The remaining free parameters are the bino and wino soft masses ($M_1$ and $M_2$), and the slepton--higgsino mass splitting $\Delta_L$ as defined in the text.}
	\label{ScanRange}
  \end{table}

\subsection{\label{sec:numerics}Numerical Results and Discussion}

\subsubsection{DM aninihilation}

\begin{figure}[htbp]
	\centering
	\includegraphics[width=1.0 \textwidth]{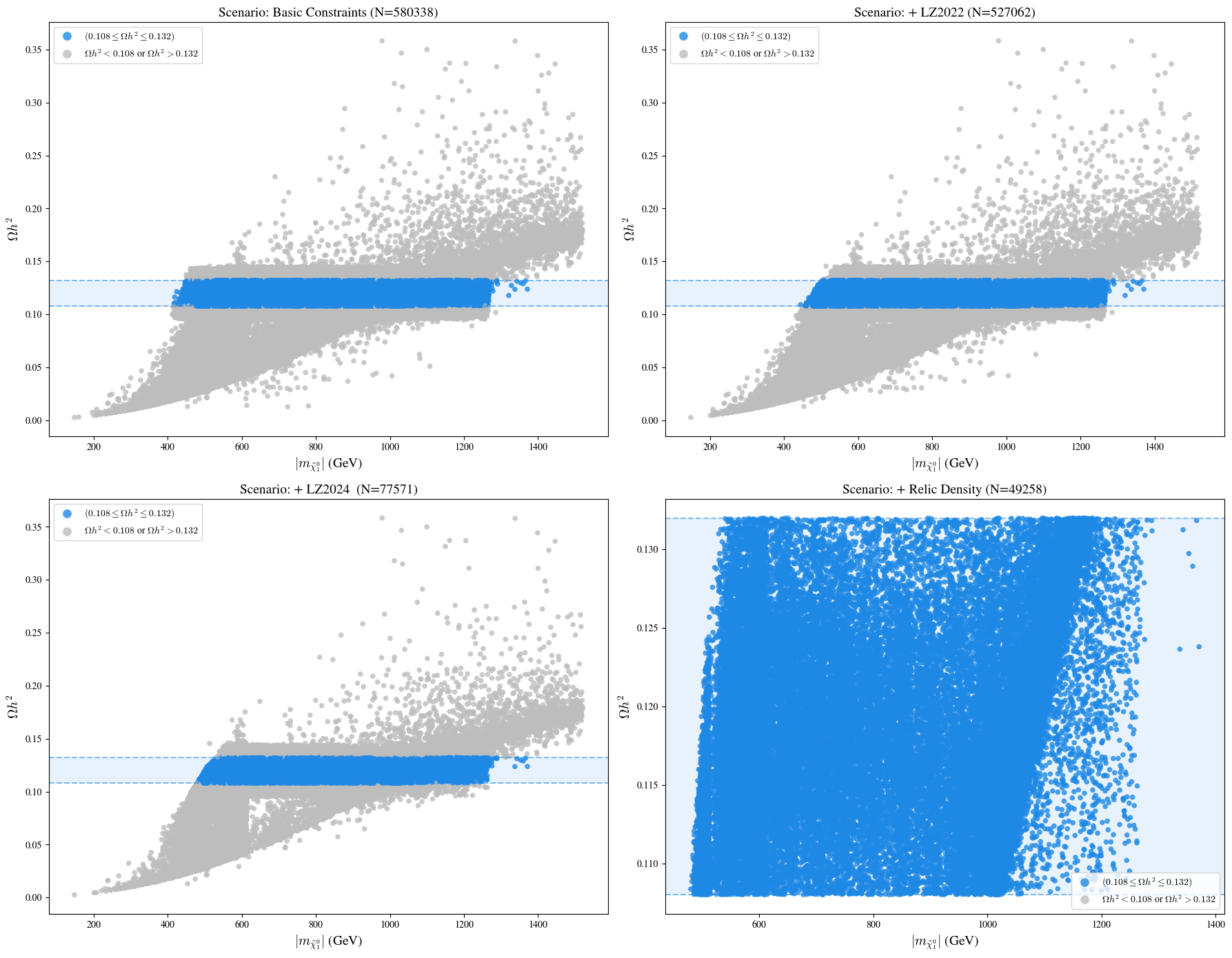} 
	\caption{The predicted relic density as a function of DM mass for four successive stages of constraints: (Top-left) after applying our baseline cuts; (Top-right) adding the LZ2022 SI constraint; (Bottom-left) further imposing the stronger LZ2024 SI constraint; (Bottom-right) additionally imposing the relic-density requirement. Note that the botom right panel is actually a zoomed-in version of the blue band for bottom-left panel.}
	\label{Figure1}
\end{figure}

Following a comprehensive scan of the parameter space, we generated a dataset containing approximately $1.1 \times 10^6$ points. In this section, we systematically evaluate these points by progressively imposing phenomenological and experimental constraints to quantify the impact of direct detection limits. Given that the LZ 2024 SI limits are significantly more stringent than the 2022 release, we explicitly compare the exclusion power of both datasets.

Figure~\ref{Figure1} illustrates the predicted relic density as a function of the dark matter mass across four successive stages of constraints, as summarized in the caption. The Baseline constraints are constraints which are illustrated in $\mathcal{L}$ constraints earlier,i.e Higgs sector constraints,Flavor Physics Constraints as well as Higgsino constraints.

The progression of the panels clearly demonstrates the evolution of the lower bound on the dark matter mass. In the absence of direct detection limits (Top-Left), the relic density requirement alone (indicated by the horizontal blue band) permits dark matter masses as low as $m_{\chi} \gtrsim 400~\mathrm{GeV}$. This deviation from the canonical $\sim 1.1~\mathrm{TeV}$ thermal higgsino limit is made possible by the presence of light, mass-degenerate sleptons, whose co-annihilations are less efficient than those of higgsinos. 

Imposing the LZ 2022 SI constraint (Top-Right) excludes a portion of the low-mass parameter space, raising the effective lower bound to slightly above $m_{\chi} \sim 450~\mathrm{GeV}$. Crucially, the more stringent LZ 2024 SI limit(Bottom-Left) further reduces the viable parameter space, shifting the lower mass bound to approximately 500 GeV. Note that in Figure~\ref{Figure1}, we focus solely on SI limits when discussing the DM mass floor, as SD constraints are substantially weaker in the higgsino-dominated regime and do not qualitatively alter the lower bound.

For completeness, the number of surviving points after each stage is annotated in the header of each subplot. Starting from the raw sample, the Baseline cuts retain 580,338 points. The application of the LZ 2022 SI limit reduces this set to 527,062 points. However, the impact of the LZ 2024 SI limit is much more severe, leaving only 77,571 points. Finally, enforcing the strict relic density requirement (Bottom-Right) leaves a final sample of 49,258 points. This represents an $\simeq 90\%$ reduction relative to the post-baseline sample, highlighting the significant tension imposed by the latest direct detection results.

\begin{figure}[htbp]
	\centering
	\includegraphics[width=1.0\textwidth]{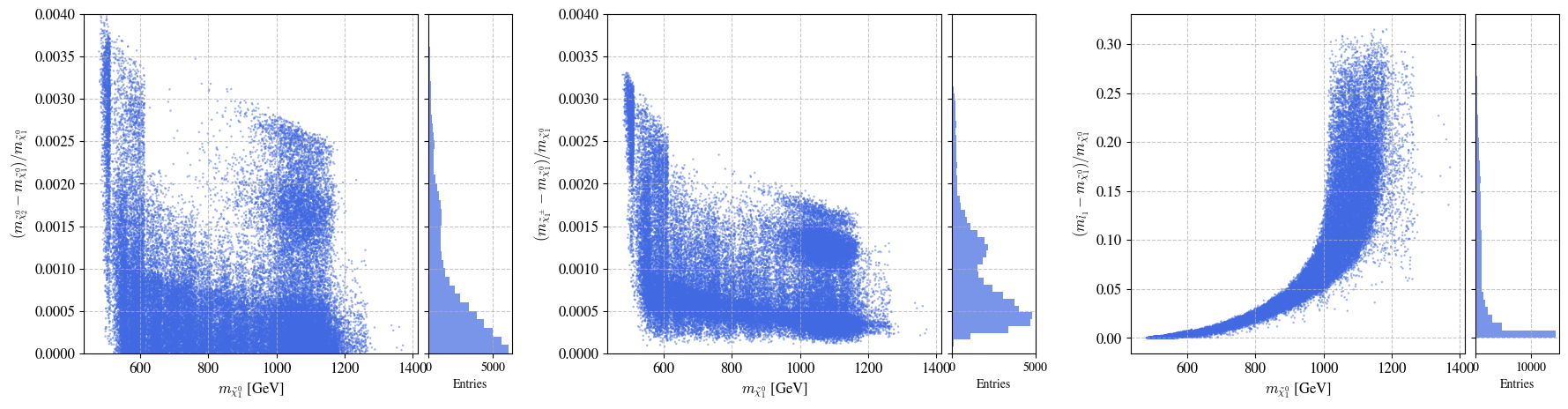} 
	\caption{The mass splittings between $\tilde{\chi}_2^0$ and $\tilde{\chi}_1^0$ (left), between $\tilde{\chi}_1^\pm$ and $\tilde{\chi}_1^0$ (middle), and between the lightest slepton-sector state $\tilde{l}_1$ and $\tilde{\chi}_1^0$ (right), shown as functions of the LSP mass for the final sample that satisfies all constraints, including the LZ2024 SI and SD limits and the relic-density requirement. For each case, the left subpanel shows the mass splitting versus the LSP mass, while the right shows the corresponding histogram of relative mass splittings. In the right panel, the label $\tilde{l}_1$ is used as a shorthand for the lightest slepton-sector coannihilating state, defined by $m_{\tilde{l}_1}\equiv \min\!\big(m_{\tilde{l}_1^\pm},\,m_{\tilde{\nu}_1}\big)$. Although multiple quasi-degenerate sleptons and sneutrinos jointly participate in the early-universe coannihilation, the mass of this lightest state effectively represents the relevant mass scale of the coannihilating sector.}
	\label{Figure2}
\end{figure}

To elucidate the physical mechanisms governing the surviving parameter space further, particularly the role of slepton co-annihilation, we examine the mass splittings between the LSP and the next-to-lightest supersymmetric particles (NLSPs), specifically $\tilde{\chi}_2^0$, $\tilde{\chi}_1^\pm$, and the lightest slepton-sector state $\tilde{l}_1$. Note that, as defined in the caption of Figure~\ref{Figure2}, the label $\tilde{l}_1$ here refers to the lightest mass eigenstate among the charged slepton $\tilde{l}_1^\pm$ and the sneutrino $\tilde{\nu}_1$. The results are presented in Figure~\ref{Figure2}.

For this analysis, we employ the final dataset that satisfies \textit{all} imposed constraints, including the stringent LZ 2024 SI and SD limits as well as the relic density requirement. While SD constraints were omitted from the progressive evolution shown in Figure~\ref{Figure1} for narrative clarity (as they do not determine the absolute lower mass bound), their inclusion here ensures a rigorous characterization of the final viable parameter space.

Figure~\ref{Figure2} displays the relative mass splitting, defined as $\delta m = (m_{\text{NLSP}} - m_{\text{LSP}})/m_{\text{LSP}}$, between the LSP $\tilde{\chi}_1^0$ and its co-annihilation partners. The figure is divided into three panels corresponding to $\tilde{\chi}_2^0$ (left), $\tilde{\chi}_1^\pm$ (middle), and the lightest slepton $\tilde{l}_1$ (right). Each panel features a scatter plot of the mass splitting versus the LSP mass, accompanied by a marginal histogram illustrating the distribution of the splitting values.

Two distinct phenomenological regimes are immediately apparent from these distributions. The higgsino sector, depicted in the left and middle panels, is characterized by the expected high degree of degeneracy; the mass splittings for $\tilde{\chi}_2^0$ and $\tilde{\chi}_1^\pm$ typically remain below 1\% across the entire dark matter mass range. In contrast, the slepton sector (right panel) exhibits a strong correlation between mass splitting and the dark matter mass. As the LSP mass decreases, the slepton mass splitting systematically drops, approaching near-exact degeneracy ($\delta m \to 0$) around 500 GeV. The sharp peaks in the histograms at low splitting values further confirm that a significant fraction of the valid points necessitate nearly degenerate sleptons.

This behaviour is directly linked to the mechanisms that set the dark matter relic abundance. 
For $m_{\tilde{\chi}_1^0} \sim 1~\mathrm{TeV}$, the correct relic density is predominantly obtained through efficient co-annihilations within the nearly degenerate higgsino multiplet itself. 
In this regime, $\tilde{\chi}_2^0$ and $\tilde{\chi}_1^\pm$ remain close in mass to the LSP, and their annihilations into electroweak gauge bosons yield a large effective cross section $\langle \sigma v \rangle_{\text{eff}}$, while sleptons are sufficiently heavier that their contribution to freeze-out is negligible.

As the dark matter mass is lowered into the sub-TeV range, the higgsino DM would typically predict an even larger $\langle \sigma v \rangle_{\text{eff}}$ and thus an under-abundance of dark matter.  
In our setup, this tendency is moderated by the presence of light sleptons that become increasingly degenerate with the LSP.  
In the parameter region selected by our scan, the slepton–slepton and slepton–higgsino (co-)annihilation cross sections are generally smaller than the dominant higgsino–higgsino channels.  
Consequently, when sleptons enter chemical equilibrium with the LSP and become nearly mass-degenerate, they increase the effective number of degrees of freedom during freeze-out without proportionally enhancing the total annihilation rate.  
This additional population of more weakly annihilating states acts to dilute the higgsino annihilation efficiency, effectively reducing $\langle \sigma v \rangle_{\text{eff}}$ relative to the pure-higgsino expectation and thereby raising the relic abundance to the observed cosmological value.  
The requirement that this dilution be sufficiently strong explains why the slepton mass splitting must shrink as $m_{\tilde{\chi}_1^0}$ approaches the $\sim 500~\mathrm{GeV}$ lower bound, as clearly seen in the right panel of Figure~\ref{Figure2}. Thus, it is the less efficient coannihilation of sleptons compared to higgsinos that reduces the effective annihilation cross section, leading to an increase in the relic density.

\subsubsection{Impact of $M_1$ and $M_2$ on Direct Detection}

\begin{figure}[htbp]
	\centering
	\resizebox{1.0\textwidth}{!}{
	\includegraphics[width=1.0\textwidth]{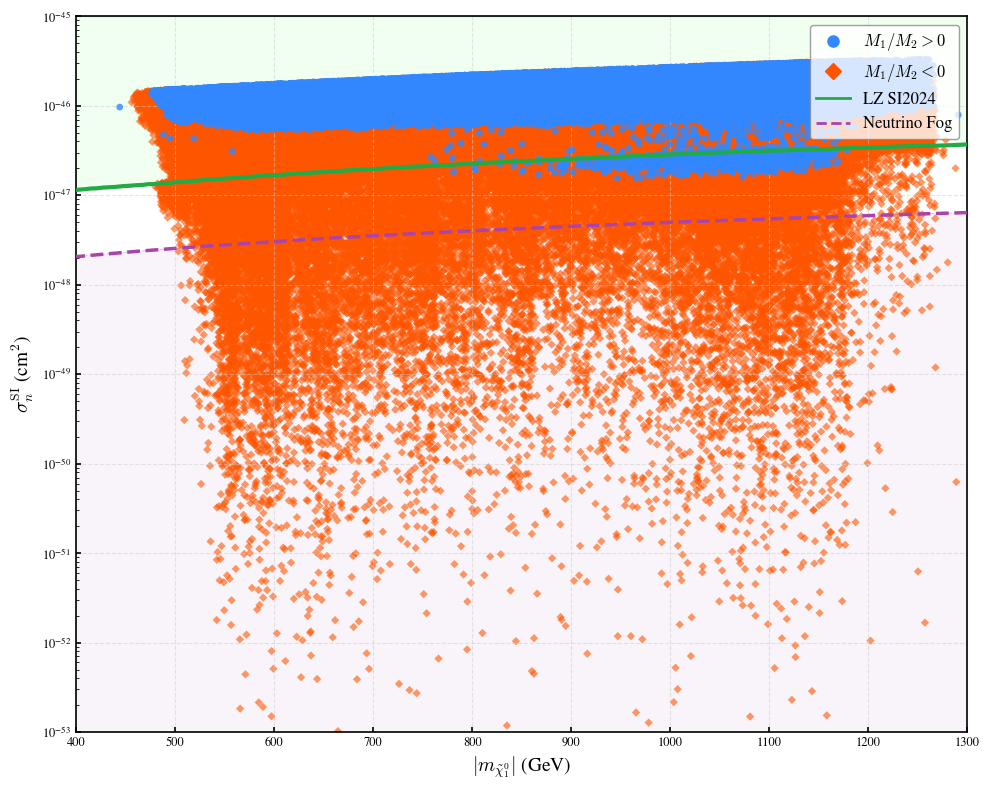}
	\includegraphics[width=1.0\textwidth]{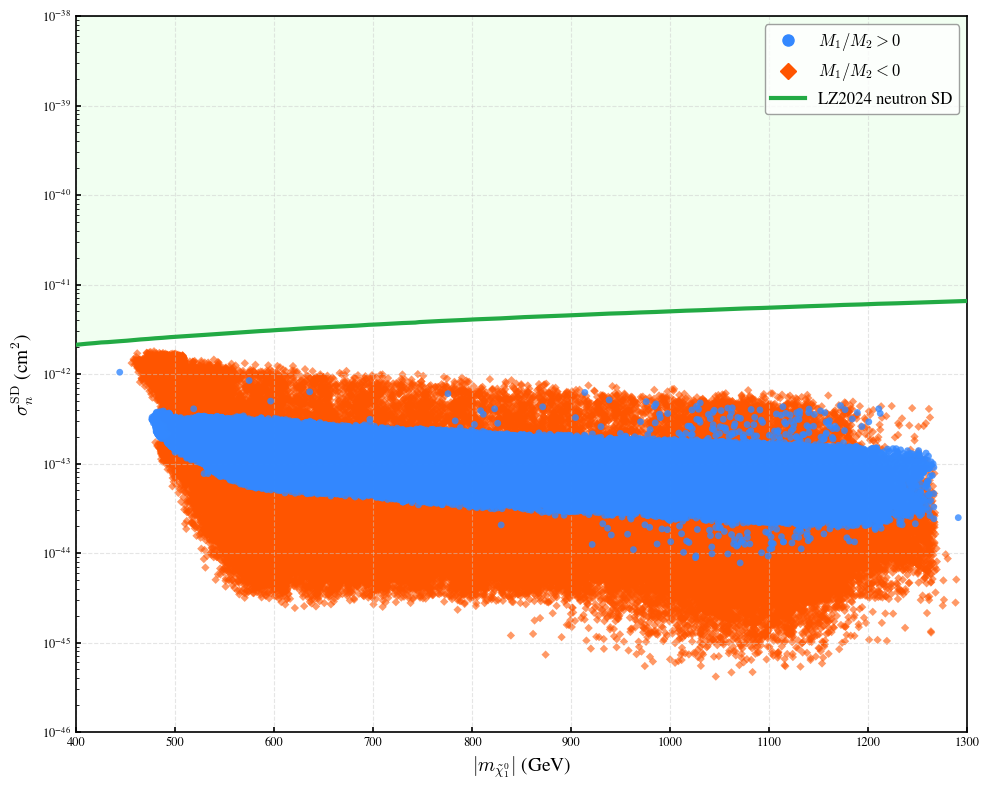}
	}
	\caption{The SI(left pannel) and neutron SD(right pannel)  cross section as a function of DM mass for samples that pass all constraints except LZ2024 SI limit. The shaded area above the green line is excluded by LZ2024 SI limit. Besides,we also plot the neutrino fog region below the dashed bright purple line for the left pannel. The orange and blue points represent the cases where $M_1$ and $M_2$ have the same sign and opposite sign respectively.
	\label{Figure3}}
\end{figure}

As previously discussed, the relative sign between the gaugino mass parameters $M_1$ and $M_2$ plays a decisive role in determining the direct detection scattering rates. To illustrate this dependence, Figure~\ref{Figure3} presents both the SI and SD nucleon cross-sections for the parameter points that satisfy all prior constraints, including the LZ 2022 limits. The solid green lines denote the latest exclusion limits from the LZ 2024 experiment, allowing for a critical evaluation of their impact on scenarios with same-sign ($M_1/M_2 > 0$, blue circles) and opposite-sign ($M_1/M_2 < 0$, orange diamonds) gaugino masses.

In the left panel, the dichotomy in exclusion power between the two scenarios is striking. For the same-sign case(blue circles), the constructive interference in the LSP-Higgs coupling (Eq.~\ref{eq:DM_HiggsCoupling}) leads to generally enhanced scattering rates, which are now severely constrained by LZ 2024. This is particularly evident for dark matter masses below 800~GeV, where $M_1$ and $M_2$ are relatively light, resulting in significant gaugino-higgsino mixing and large cross-sections that are robustly excluded. However, a distinct cluster of points survives around 1~TeV. In this regime, the heavy mass scale of the gauginos ($M_{1,2} \gg \mu$) suppresses their admixture in the LSP state. Since the gaugino components enter the effective coupling inversely proportional to their masses, the scattering cross-section is naturally driven down. Consequently, these points approach the pure higgsino limit, where the suppressed coupling allows them to evade even the stringent LZ 2024 constraints despite the lack of destructive interference.

In sharp contrast, the opposite-sign case(orange diamonds) remains largely unconstrained by the new limits. This resilience is a direct consequence of destructive interference within the coupling structure, where the scattering amplitude can be suppressed by orders of magnitude regardless of the specific mass scale. A significant number of these points populate the region well below the current limit, with some extending down to the neutrino fog(indicated by the dashed purple line)—an irreducible background that poses a fundamental challenge for future direct detection experiments\cite{OHare:2021utq,Bloch:2024suj}.

The right panel illustrates the SD neutron cross-section for the same dataset. Evidently, the current LZ 2024 SD limit is orders of magnitude weaker than the predicted rates and thus provides no effective constraints on this parameter space at present. Interestingly, however, the theoretical pattern mirrors that of the SI case: the same-sign scenario(blue circles) generally yields larger SD cross-sections than the opposite-sign counterpart(orange diamonds). This correlation exists because the $Z$-boson coupling (Eq.~\ref{eq:DM_ZCoupling}), which mediates the SD interaction, depends on the neutralino mixing matrices in a manner that exhibits analogous constructive or destructive tendencies governed by the relative signs of the gaugino masses.

\begin{figure}[htbp]
	\centering
	\includegraphics[width=1.0\textwidth]{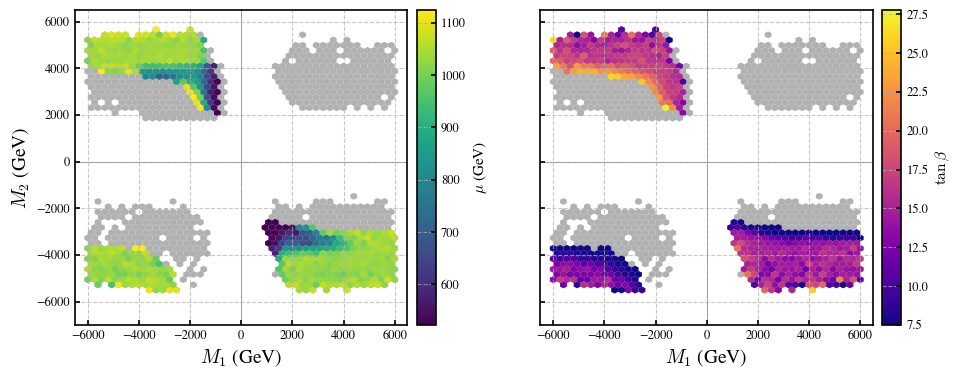}
	\caption{Distribution of the surviving parameter points in the $M_1$-$M_2$ plane, visualized using hexagonal binning. 
	The colored bins represent points that satisfy all constraints, including the latest LZ2024 results. The color scale indicates the mean value of $\mu$ (left panel) and $\tan\beta$ (right panel) within each bin. Grey bins denote points that were allowed by previous constraints (including LZ2022) but are now excluded by the LZ2024 results.\label{Figure4}}
\end{figure}

To further illustrate the impact of the LZ2024 constraints, Figure~\ref{Figure4} shows the distribution of our sampled parameter points in the $(M_1, M_2)$ gaugino mass plane using hexagonal binning. Grey bins denote points that satisfy all constraints including the LZ2022 limits but are excluded once the LZ2024 results are imposed. Colored bins represent points that remain viable after applying the LZ2024 limits; the color scale encodes the mean value of the higgsino mass parameter $\mu$ in the left panel and of $\tan\beta$ in the right panel within each bin.

Comparing the grey and colored regions reveals a dramatic and highly asymmetric reshaping of the viable parameter space. Before imposing LZ2024, points with relatively light gauginos ($|M_1|, |M_2| \sim 2000~\text{GeV}$ or even lower) were allowed across all four quadrants. However, the new constraints impose a severe penalty on regions where $M_1$ and $M_2$ have the same sign. The first quadrant($M_1, M_2 > 0$) is completely emptied, and in the third quadrant ($M_1, M_2 < 0$), the surviving points are pushed to very high masses ($|M_2| \gtrsim 4000~\text{GeV}$). In stark contrast, the opposite-sign quadrants ($M_1/M_2 < 0$) retain a much larger portion of the parameter space. While the lower bound on $|M_2|$ is lifted moderately(shifting towards $3000\text{--}4000~\text{GeV}$ as $|M_1|$ increases), light states with $|M_1| \sim 2000~\text{GeV}$ and moderate $|M_2|$ remain viable.

This pronounced difference between the same-sign and opposite-sign regions points directly to the structure of the spin-independent DM–Higgs coupling. For a higgsino-like LSP this coupling contains two leading terms controlled by $M_1$ and $M_2$, often associated with bino- and wino-like components. When $M_1$ and $M_2$ share the same sign, these $M_1$- and $M_2$-dependent terms add constructively, as seen in Eq.~\ref{eq:DM_HiggsCoupling}, leading to an enhanced SI cross-section that is now strongly ruled out by LZ2024.

Despite this overall trend, an asymmetry is visible between the two same-sign quadrants. The first quadrant is entirely removed by the new bound, whereas a small subset of points survives in the third quadrant. The surviving region is characterized by large gaugino masses, $|M_1|, |M_2| \gtrsim 2000~\text{GeV}$, higgsino mass parameters $|\mu|$ of order $1000~\text{GeV}$ that remain well below $|M_{1,2}|$, and comparatively small values of $\tan\beta$ within our scan range. The large $|M_{1,2}|$ suppresses the higgsino–gaugino mixing angles, leading to decoupling, while the smaller $\tan\beta$ values enhance $\sin 2\beta$, moving the coupling closer to the spin-independent blind-spot condition identified in Ref.~\cite{Cheung:2012qy}, although it does not enforce an exact cancellation in our parameter range.

In contrast, the second and fourth quadrants($M_1/M_2 < 0$) remain largely unconstrained. Here the $M_1$- and $M_2$-dependent terms in the DM–Higgs coupling enter with opposite signs and partially cancel, which naturally suppresses the SI cross-section. This cancellation mechanism allows points with comparatively smaller gaugino masses $|M_1|, |M_2| \sim 2000 \text{--}3000~\text{GeV}$, to remain viable after LZ2024. The left panel further shows that these surviving opposite-sign points with lower $|M_1|$ and $|M_2|$ tend to have smaller values of $\mu$ as well. This correlation is a direct consequence of our scan setup, which assumes $|\mu| \ll |M_1|, |M_2|$ in order to realize a higgsino-dominated LSP.

Overall, in the opposite-sign region $M_1/M_2 < 0$, the partial cancellation between the $M_1$- and $M_2$-dependent pieces of the DM–Higgs coupling leads to a pronounced suppression of the SI cross section. This pattern is qualitatively analogous to the direct-detection blind spots discussed in Ref.~\cite{Cheung:2012qy}, although our surviving points do not lie on the exact blind-spot surfaces, which would require, for example, $M_{1,2} + \mu \sin 2\beta = 0$ or $\tan\beta = 1$. Instead, they are located in a decoupling regime with $|M_{1,2}| \gg |\mu|$ and moderate-to-large $\tan\beta$, where higgsino purity and partial cancellation between the $M_1$ and $M_2$ terms jointly drive the SI cross section below the LZ2024 limit.

\section{Benchmark Points\label{BMP}}

To provide a concrete illustration of the phenomenology discussed above, we present four benchmark points in Table~\ref{BMP12} and Table~\ref{BMP34}, corresponding to the first through fourth quadrants of Figure~\ref{Figure4} respectively. Together they exemplify the different mechanisms that allow higgsino-like DM to satisfy the relic-density and direct-detection constraints in the same-sign and opposite-sign regions of $(M_1,M_2)$.

As shown in the tables, all four benchmarks feature a higgsino-dominated DM candidate, evidenced by the nearly mass-degenerate spectrum of $\tilde{\chi}_{1,2}^0$ and $\tilde{\chi}_1^\pm$ with masses close to the $\mu$ parameter. 
Point P1, characterized by positive $M_1$ and $M_2$ and a relatively light DM state, was permitted by LZ2022 but is now excluded by LZ2024. 
This behaviour directly reflects the discussion in Section~\ref{sec:numerics}: in the first quadrant the $M_1$- and $M_2$-dependent pieces of the DM--Higgs coupling add constructively, leading to a comparatively large SI cross section that has now been probed by the improved sensitivity of LZ2024.

In contrast, Point P3 lies in the same-sign region with large negative $M_1$ and $M_2$ and remains viable thanks to the decoupling of heavy gauginos. For P3, the DM mass is approximately $850.2~\text{GeV}$. 
Although the mass splitting between the lightest slepton and the DM is about $30~\text{GeV}$, the much smaller splittings ($\sim 2~\text{GeV}$) with the other light higgsinos ensure that co-annihilation with $\tilde{\chi}_1^\pm$ and $\tilde{\chi}_2^0$ is the dominant mechanism for achieving the correct relic density. 
At the same time, the large $|M_{1,2}|$ values suppress the higgsino--gaugino mixing angles, and the relatively small $\tan\beta$ enhances $\sin 2\beta$, placing P3 in the decoupling regime identified in Figure~\ref{Figure4} and moving it close to the spin-independent blind-spot condition of Ref.~\cite{Cheung:2012qy}. 
As a result, the effective DM--Higgs coupling, and hence the SI cross section, is strongly reduced.

Points P2 and P4 represent scenarios where $M_1$ and $M_2$ have opposite signs, corresponding to the cancellation regions in Figure~\ref{Figure4}. 
In both cases, the mass splittings between the DM and the other higgsinos are extremely small ($\lesssim 1~\text{GeV}$), making higgsino co-annihilation highly efficient. 
However, as the DM mass decreases from P2 ($\sim 602~\text{GeV}$) to P4 ($\sim 498~\text{GeV}$), the mass degeneracy with the lightest slepton becomes more pronounced.
Consequently, the contribution from slepton co-annihilation channels increases from approximately $1.0\%$ in P2 to about $9.5\%$ in P4, while higgsino co-annihilation remains the dominant process in both cases.
From the point of view of direct detection, the opposite signs of $M_1$ and $M_2$ imply that the $M_1$- and $M_2$-dependent terms in the DM--Higgs coupling partially cancel, in line with the behaviour seen in the second and fourth quadrants of Figure~\ref{Figure4}. 
This destructive interference suppresses the SI cross section even though the gaugino masses in P2 and P4 are smaller than in P3, illustrating the complementarity between decoupling and cancellation effects.

Overall, these benchmark points demonstrate a clear trend: as the higgsino-like DM mass decreases, co-annihilation with sleptons becomes increasingly important, although it still provides only a subdominant contribution compared to higgsino co-annihilation in our benchmarks. 
At the same time, they explicitly realize the three main mechanisms identified in Section~\ref{sec:numerics}: (i) constructive interference in the same-sign region leading to exclusion by LZ2024 (P1), (ii) suppression of the DM--Higgs coupling via gaugino decoupling and proximity to a blind spot (P3), and (iii) partial cancellation of the $M_1$ and $M_2$ contributions when $M_1/M_2<0$ (P2 and P4). 
This behaviour corroborates our numerical analysis in Section~\ref{sec:numerics} and the trends shown in Figures~\ref{Figure2} and~\ref{Figure4}.

In addition, we have checked all four benchmark points against the latest LHC 13~TeV electroweakino searches. 
The relevant production channels are 
$p p \to \tilde{\chi}_i^\pm \tilde{\chi}_j^\mp$, 
$p p \to \tilde{\chi}_i^0 \tilde{\chi}_j^0$, 
and 
$p p \to \tilde{\chi}_i^\pm \tilde{\chi}_j^0$ 
with $i,j=1,2$. 
We generate parton-level events with MadGraph5\_aMC@NLO 2.9~\cite{alwall2011madgraph,alwall2014automated}, perform parton shower and hadronization using Pythia 8.3~\cite{Bierlich:2022pfr,2006pythia}, and simulate the detector response with Delphes-3.4~\cite{de2014delphes}. 
The resulting detector-level events are then passed to CheckMATE2 2.0.34~\cite{Dercks:2016npn}, which compares our signal predictions with the published ATLAS and CMS analyses and determines whether a given parameter point is excluded at 95\%~CL. For each benchmark we evaluate the CheckMATE variable $R = \max_i (S_i / S_i^{95})$,where $S_i$ is the predicted signal yield in signal region $i$ and $S_i^{95}$ is the corresponding experimental upper limit. We find that the largest $R$ value among our benchmarks is about $0.02$, well below the exclusion threshold $R=1$, so none of the points is currently ruled out by existing LHC searches.
However, the High-Luminosity LHC(HL-LHC) and High Energy LHC(HE-LHC)\cite{Aboubrahim:2021xfi}, as well as future indirect searches\cite{ackermann2015searching,Ackermann:2015zua}(e.g.\ with H.E.S.S. and future gamma-ray telescopes), will substantially improve the sensitivity to higgsino-like electroweakinos and could probe large parts of the parameter space illustrated by our benchmark points~\cite{Yue:2025dqe}.

\begin{table}[t]
  \centering
  \scriptsize
  \setlength{\tabcolsep}{5pt}
  \begin{tabular}{l l l l | l l l l}
    \toprule
    \multicolumn{4}{c}{\textbf{Point P1}} & \multicolumn{4}{c}{\textbf{Point P2}} \\
    \midrule
    \multicolumn{2}{l}{Input Parameter} & \multicolumn{2}{l|}{Value} &
    \multicolumn{2}{l}{Input Parameter} & \multicolumn{2}{l}{Value} \\
    \midrule
    \multicolumn{2}{l}{$\tan\beta,\Delta_L$} & \multicolumn{2}{l|}{18.68,0.14} &
    \multicolumn{2}{l}{$\tan\beta,\Delta_L$} & \multicolumn{2}{l}{27.06,0.18} \\
    \multicolumn{2}{l}{$\mu,M_1,M_2$} & \multicolumn{2}{l|}{483.90,2470.27,2665.33} &
    \multicolumn{2}{l}{$\mu,M_1,M_2$} & \multicolumn{2}{l}{588.40,-1195.71,3337.25} \\
    \multicolumn{2}{l}{$m_A,A_t$} & \multicolumn{2}{l|}{1254.38,4517.42} &
    \multicolumn{2}{l}{$m_A,A_t$} & \multicolumn{2}{l}{1702.39,2817.50} \\
    \midrule
    \multicolumn{2}{l}{Particle} & \multicolumn{2}{l|}{Mass Spectrum (Pole)} &
    \multicolumn{2}{l}{Particle} & \multicolumn{2}{l}{Mass Spectrum (Pole)} \\
    \multicolumn{2}{l}{$\tilde{\chi}^0_1,\tilde{\chi}^0_2,\tilde{\chi}^0_3,\tilde{\chi}^0_4$} & \multicolumn{2}{l|}{494.6,-497.0,2465.2,2642.8} &
    \multicolumn{2}{l}{$\tilde{\chi}^0_1,\tilde{\chi}^0_2,\tilde{\chi}^0_3,\tilde{\chi}^0_4$} & \multicolumn{2}{l}{-601.7,602.4,-1191.8,3296.9} \\
    \multicolumn{2}{l}{$\tilde{\chi}_{1}^{\pm},\tilde{\chi}_{2}^{\pm}$} & \multicolumn{2}{l|}{496.0,2642.9} &
    \multicolumn{2}{l}{$\tilde{\chi}_{1}^{\pm},\tilde{\chi}_{2}^{\pm}$} & \multicolumn{2}{l}{601.7,3296.9} \\
    \multicolumn{2}{l}{$h,H,A$} & \multicolumn{2}{l|}{124.8,1684.1,1684.1} &
    \multicolumn{2}{l}{$h,H,A$} & \multicolumn{2}{l}{125.0,2178.6,2178.6} \\
    \multicolumn{2}{l}{$\tilde{l}_1^{\pm},\tilde{l}_2^{\pm},\tilde{l}_3^{\pm}$} & \multicolumn{2}{l|}{494.73,501.14,501.17} &
    \multicolumn{2}{l}{$\tilde{l}_1^{\pm},\tilde{l}_2^{\pm},\tilde{l}_3^{\pm}$} & \multicolumn{2}{l}{603.99,610.01,610.03} \\
    \multicolumn{2}{l}{$\tilde{l}_4^{\pm},\tilde{l}_5^{\pm},\tilde{l}_6^{\pm}$} & \multicolumn{2}{l|}{538.46,538.48,543.22} &
    \multicolumn{2}{l}{$\tilde{l}_4^{\pm},\tilde{l}_5^{\pm},\tilde{l}_6^{\pm}$} & \multicolumn{2}{l}{700.32,700.33,704.27} \\
    \multicolumn{2}{l}{$\tilde{\nu}_1,\tilde{\nu}_2,\tilde{\nu}_3$} & \multicolumn{2}{l|}{494.71,495.10,495.11} &
    \multicolumn{2}{l}{$\tilde{\nu}_1,\tilde{\nu}_2,\tilde{\nu}_3$} & \multicolumn{2}{l}{604.59,604.91,604.91} \\
    \midrule
    \multicolumn{2}{l}{Mixing Matrix ($N_{ij}$)} & \multicolumn{2}{l|}{Element Value [\%]} &
    \multicolumn{2}{l}{Mixing Matrix ($N_{ij}$)} & \multicolumn{2}{l}{Element Value [\%]} \\
    \multicolumn{2}{l}{$N_{11},N_{12},N_{13},N_{14}$} & \multicolumn{2}{l|}{-1.7,2.6,-70.8,70.6} &
    \multicolumn{2}{l}{$N_{11},N_{12},N_{13},N_{14}$} & \multicolumn{2}{l}{5.1,1.3,70.7,70.5} \\
    \multicolumn{2}{l}{$N_{21},N_{22},N_{23},N_{24}$} & \multicolumn{2}{l|}{-1.0,1.6,70.6,70.8} &
    \multicolumn{2}{l}{$N_{21},N_{22},N_{23},N_{24}$} & \multicolumn{2}{l}{-1.8,-2.1,70.7,-70.7} \\
    \multicolumn{2}{l}{$N_{31},N_{32},N_{33},N_{34}$} & \multicolumn{2}{l|}{100.0,1.0,-0.5,1.8} &
    \multicolumn{2}{l}{$N_{31},N_{32},N_{33},N_{34}$} & \multicolumn{2}{l}{99.9,-0.1,-2.3,-4.9} \\
    \multicolumn{2}{l}{$N_{41},N_{42},N_{43},N_{44}$} & \multicolumn{2}{l|}{0.9,-100.0,-0.7,3.1} &
    \multicolumn{2}{l}{$N_{41},N_{42},N_{43},N_{44}$} & \multicolumn{2}{l}{0,-100,-0.5,2.4} \\
    \midrule
    \multicolumn{2}{l}{Slepton Mixing ($U_{ij}$)} & \multicolumn{2}{l|}{Element Value [\%]} &
    \multicolumn{2}{l}{Slepton Mixing ($U_{ij}$)} & \multicolumn{2}{l}{Element Value [\%]} \\
    \multicolumn{2}{l}{$U_{13},U_{16},U_{22},U_{25}$} & \multicolumn{2}{l|}{93,34,-99,-2.5} &
    \multicolumn{2}{l}{$U_{13},U_{16},U_{22},U_{25}$} & \multicolumn{2}{l}{97,23,-99,-1.5} \\
    \multicolumn{2}{l}{$U_{31},U_{34},U_{41},U_{44}$} & \multicolumn{2}{l|}{-99,-0.01,0.01,-99} &
    \multicolumn{2}{l}{$U_{31},U_{34},U_{41},U_{44}$} & \multicolumn{2}{l}{99,0.007,0.007,99} \\
    \multicolumn{2}{l}{$U_{52},U_{55},U_{63},U_{66}$} & \multicolumn{2}{l|}{-2.5,99,-34,94} &
    \multicolumn{2}{l}{$U_{52},U_{55},U_{63},U_{66}$} & \multicolumn{2}{l}{1.5,-99,-23,97} \\
    \midrule
    \multicolumn{3}{l}{Primary annihilation ($>1\%$)} & Fraction [\%] &
    \multicolumn{3}{l}{Primary annihilation ($>1\%$)} & Fraction [\%] \\
    \multicolumn{3}{l}{$\tilde{\chi}^0_1\,\tilde{\chi}_{1}^{-}\rightarrow d\bar{u}/s\bar{c}/b\bar{t}/\bar{\nu}_e e/\bar{\nu}_\mu \mu/\bar{\nu}_\tau \tau$} & 22.4 &
    \multicolumn{3}{l}{$\tilde{\chi}^0_1\,\tilde{\chi}_{1}^{-}\rightarrow d\bar{u}/s\bar{c}/b\bar{t}/\bar{\nu}_e e/\bar{\nu}_\mu \mu/\bar{\nu}_\tau \tau$} & 23.1 \\
    \multicolumn{3}{l}{$\tilde{\chi}^0_2\,\tilde{\chi}_{1}^{-}\rightarrow d\bar{u}/s\bar{c}/b\bar{t}/\bar{\nu}_e e/\bar{\nu}_\mu \mu/\bar{\nu}_\tau \tau$} & 14.8 &
    \multicolumn{3}{l}{$\tilde{\chi}^0_1\,\tilde{\chi}_{1}^{-}\rightarrow \gamma W_{1}^{-}$} & 1.0 \\
    \multicolumn{3}{l}{$\tilde{\chi}^0_1\,\tilde{\chi}^0_1\rightarrow W^{-}W^{+}/ZZ$} & 3.9 &
    \multicolumn{3}{l}{$\tilde{\chi}^0_2\,\tilde{\chi}_{1}^{-}\rightarrow d\bar{u}/s\bar{c}/b\bar{t}/\bar{\nu}_e e/\bar{\nu}_\mu \mu/\bar{\nu}_\tau \tau$} & 18.9 \\
    \multicolumn{3}{l}{$\tilde{\chi}_{1}^{-}\,\tilde{\chi}_{1}^{+}\rightarrow u\bar{u}/c\bar{c}/t\bar{t}/d\bar{d}/s\bar{s}/W^{-}W^{+}$} & 8.9 &
    \multicolumn{3}{l}{$\tilde{\chi}_{1}^{-}\,\tilde{\chi}_{1}^{+}\rightarrow u\bar{u}/c\bar{c}/t\bar{t}/d\bar{d}/s\bar{s}/W^{-}W^{+}$} & 9.4 \\
    \multicolumn{3}{l}{$\tilde{\chi}^0_1\,\tilde{\chi}^0_2\rightarrow d\bar{d}/s\bar{s}/u\bar{u}/c\bar{c}/b\bar{b}$} & 7.5 &
    \multicolumn{3}{l}{$\tilde{\chi}^0_1\,\tilde{\chi}^0_1\rightarrow W^{-}W^{+}/ZZ$} & 5.3 \\
    \multicolumn{3}{l}{$\tilde{\nu}_1\,\tilde{\nu}_1^{*}\rightarrow W^{-}W^{+}/ZZ$} & 2.8 &
    \multicolumn{3}{l}{$\tilde{\chi}^0_1\,\tilde{\chi}^0_2\rightarrow u\bar{u}/c\bar{c}/d\bar{d}/s\bar{s}/b\bar{b}$} & 10.6 \\
    \multicolumn{3}{l}{$\tilde{l}_1^{+}\,\tilde{l}_1^{-}\rightarrow W^{-}W^{+}$} & 1.3 &
    \multicolumn{3}{l}{$\tilde{\chi}^0_1\,\tilde{\chi}^0_2\rightarrow \nu_e\bar{\nu}_e/\nu_\mu\bar{\nu}_\mu/\nu_\tau\bar{\nu}_\tau$} & 3.2 \\
    \multicolumn{3}{l}{$\tilde{\nu}_2\,\tilde{\nu}_2^{*}\rightarrow W^{-}W^{+}$} & 1.1 &
    \multicolumn{3}{l}{$\tilde{l}_1^{-}\,\tilde{\chi}_{1}^{-}\rightarrow W^{-}\tau^{-}$} & 1.0 \\
    \multicolumn{3}{l}{$\tilde{\nu}_3\,\tilde{\nu}_3^{*}\rightarrow W^{-}W^{+}$} & 1.1 & & \\
    \multicolumn{3}{l}{$\tilde{\nu}_1\,\tilde{l}_1^{+}\rightarrow AW^{+}$} & 1.1 & & \\
    \midrule
    \multicolumn{2}{l}{DM observable} & \multicolumn{2}{l|}{Value} &
    \multicolumn{2}{l}{DM observable} & \multicolumn{2}{l}{Value} \\
    \multicolumn{2}{l}{$\Omega h^2$} & \multicolumn{2}{l|}{0.118} &
    \multicolumn{2}{l}{$\Omega h^2$} & \multicolumn{2}{l}{0.120} \\
    \multicolumn{2}{l}{$\sigma^{\text{SI}}_{p,n}/(10^{-46}\text{cm}^2)$} & \multicolumn{2}{l|}{$1.40,1.44$} &
    \multicolumn{2}{l}{$\sigma^{\text{SI}}_{p,n}/(10^{-47}\text{cm}^2)$} & \multicolumn{2}{l}{$1.24,1.26$} \\
    \multicolumn{2}{l}{$\sigma^{\text{SD}}_{p,n}/(10^{-43}\text{cm}^2)$} & \multicolumn{2}{l|}{$3.39,2.60$} &
    \multicolumn{2}{l}{$\sigma^{\text{SD}}_{p,n}/(10^{-43}\text{cm}^2)$} & \multicolumn{2}{l}{$1.44,1.10$} \\
    \bottomrule
  \end{tabular}
  \caption{Summary of input parameters, pole mass spectra, mixing matrices, dominant (co)annihilation channels and selected dark matter observables for benchmark points P1 and P2. Point P1 is selected from the first quadrant ($M_1,M_2>0$) of Figure~\ref{Figure4}, representing a region excluded by recent LZ2024 limits, whereas P2 is chosen from the second quadrant ($M_1<0,M_2>0$) and remains a viable scenario. All dimensionful parameters are given in GeV. Input parameters, such as $m_A\equiv m_A^{\overline{\mathrm{DR}}}(Q=1\,\text{TeV})$, are running quantities, while the spectrum lists loop-corrected pole masses. For fermions, the table shows signed eigenvalues, whose absolute values correspond to the physical masses. The neutralino ($N_{ij}$) and slepton ($U_{ij}$) mixing matrix elements are listed in units of $10^{-2}$. The slepton mixing is defined in the basis $(\tilde{e}_L,\tilde{\mu}_L,\tilde{\tau}_L,\tilde{e}_R,\tilde{\mu}_R,\tilde{\tau}_R)$. For brevity, only numerically relevant entries are shown, as all others are negligible.}
  \label{BMP12}
\end{table}

\begin{table}[t]
  \centering
  \scriptsize
  \setlength{\tabcolsep}{5pt}
  \begin{tabular}{l l l l | l l l l}
    \toprule
    \multicolumn{4}{c}{\textbf{Point P3}} & \multicolumn{4}{c}{\textbf{Point P4}} \\
    \midrule
    \multicolumn{2}{l}{Input Parameter} & \multicolumn{2}{l|}{Value} &
    \multicolumn{2}{l}{Input Parameter} & \multicolumn{2}{l}{Value} \\
    \midrule
    \multicolumn{2}{l}{$\tan\beta,\Delta_L$} & \multicolumn{2}{l|}{8.28,0.28} &
    \multicolumn{2}{l}{$\tan\beta,\Delta_L$} & \multicolumn{2}{l}{27.05,0.17} \\
    \multicolumn{2}{l}{$\mu,M_1,M_2$} & \multicolumn{2}{l|}{833.09,-4833.66,-4604.87} &
    \multicolumn{2}{l}{$\mu,M_1,M_2$} & \multicolumn{2}{l}{485.86,1330.40,-2926.88} \\
    \multicolumn{2}{l}{$m_A,A_t$} & \multicolumn{2}{l|}{1147.13,3753.44} &
    \multicolumn{2}{l}{$m_A,A_t$} & \multicolumn{2}{l}{1824.18,2831.66} \\
    \midrule
    \multicolumn{2}{l}{Particle} & \multicolumn{2}{l|}{Mass Spectrum (Pole)} &
    \multicolumn{2}{l}{Particle} & \multicolumn{2}{l}{Mass Spectrum (Pole)} \\
    \multicolumn{2}{l}{$\tilde{\chi}^0_1,\tilde{\chi}^0_2,\tilde{\chi}^0_3,\tilde{\chi}^0_4$} & \multicolumn{2}{l|}{-850.2,852.6,-4551.2,-4868.2} &
    \multicolumn{2}{l}{$\tilde{\chi}^0_1,\tilde{\chi}^0_2,\tilde{\chi}^0_3,\tilde{\chi}^0_4$} & \multicolumn{2}{l}{-498.1,498.9,1326.1,-2896.9} \\
    \multicolumn{2}{l}{$\tilde{\chi}_{1}^{\pm},\tilde{\chi}_{2}^{\pm}$} & \multicolumn{2}{l|}{851.9,4551.3} &
    \multicolumn{2}{l}{$\tilde{\chi}_{1}^{\pm},\tilde{\chi}_{2}^{\pm}$} & \multicolumn{2}{l}{499.2,2897.0} \\
    \multicolumn{2}{l}{$h,H,A$} & \multicolumn{2}{l|}{125.4,1151.1,1165.1} &
    \multicolumn{2}{l}{$h,H,A$} & \multicolumn{2}{l}{125.2,1991.8,1991.7} \\
    \multicolumn{2}{l}{$\tilde{l}_1^{\pm},\tilde{l}_2^{\pm},\tilde{l}_3^{\pm}$} & \multicolumn{2}{l|}{880.98,881.442,881.443} &
    \multicolumn{2}{l}{$\tilde{l}_1^{\pm},\tilde{l}_2^{\pm},\tilde{l}_3^{\pm}$} & \multicolumn{2}{l}{498.32,504.40,504.43} \\
    \multicolumn{2}{l}{$\tilde{l}_4^{\pm},\tilde{l}_5^{\pm},\tilde{l}_6^{\pm}$} & \multicolumn{2}{l|}{989.691,989.692,989.93} &
    \multicolumn{2}{l}{$\tilde{l}_4^{\pm},\tilde{l}_5^{\pm},\tilde{l}_6^{\pm}$} & \multicolumn{2}{l}{573.03,573.05,577.95} \\
    \multicolumn{2}{l}{$\tilde{\nu}_1,\tilde{\nu}_2,\tilde{\nu}_3$} & \multicolumn{2}{l|}{877.99,878.04,878.04} &
    \multicolumn{2}{l}{$\tilde{\nu}_1,\tilde{\nu}_2,\tilde{\nu}_3$} & \multicolumn{2}{l}{498.21,498.23,498.23} \\
    \midrule
    \multicolumn{2}{l}{Neutralino Mixing ($N_{ij}$)} & \multicolumn{2}{l|}{Element Value [\%]} &
    \multicolumn{2}{l}{Neutralino Mixing ($N_{ij}$)} & \multicolumn{2}{l}{Element Value [\%]} \\
    \multicolumn{2}{l}{$N_{11},N_{12},N_{13},N_{14}$} & \multicolumn{2}{l|}{-0.7,1.3,-70.8,-70.6} &
    \multicolumn{2}{l}{$N_{11},N_{12},N_{13},N_{14}$} & \multicolumn{2}{l}{1.6,2.2,-70.8,-70.6} \\
    \multicolumn{2}{l}{$N_{21},N_{22},N_{23},N_{24}$} & \multicolumn{2}{l|}{0.6,-1.2,-70.7,70.8} &
    \multicolumn{2}{l}{$N_{21},N_{22},N_{23},N_{24}$} & \multicolumn{2}{l}{3.9,1.7,70.7,-70.6} \\
    \multicolumn{2}{l}{$N_{31},N_{32},N_{33},N_{34}$} & \multicolumn{2}{l|}{0.2,100.0,0.1,1.8} &
    \multicolumn{2}{l}{$N_{31},N_{32},N_{33},N_{34}$} & \multicolumn{2}{l}{99.9,-0.1,-1.6,3.9} \\
    \multicolumn{2}{l}{$N_{41},N_{42},N_{43},N_{44}$} & \multicolumn{2}{l|}{-100.0,0.2,0.1,0.9} &
    \multicolumn{2}{l}{$N_{41},N_{42},N_{43},N_{44}$} & \multicolumn{2}{l}{0,100.0,-0.4,-2.8} \\
    \midrule
    \multicolumn{2}{l}{Slepton Mixing ($U_{ij}$)} & \multicolumn{2}{l|}{Element Value [\%]} &
    \multicolumn{2}{l}{Slepton Mixing ($U_{ij}$)} & \multicolumn{2}{l}{Element Value [\%]} \\
    \multicolumn{2}{l}{$U_{13},U_{16},U_{22},U_{25}$} & \multicolumn{2}{l|}{99,5.9,-99,-0.3} &
    \multicolumn{2}{l}{$U_{13},U_{16},U_{22},U_{25}$} & \multicolumn{2}{l}{96,26,-99,-1.76} \\
    \multicolumn{2}{l}{$U_{31},U_{34},U_{41},U_{44}$} & \multicolumn{2}{l|}{100,0.001,0.001,-100} &
    \multicolumn{2}{l}{$U_{31},U_{34},U_{41},U_{44}$} & \multicolumn{2}{l}{-99,-0.008,0.008,-99} \\
    \multicolumn{2}{l}{$U_{52},U_{55},U_{63},U_{66}$} & \multicolumn{2}{l|}{0.35,-99,-5.9,99} &
    \multicolumn{2}{l}{$U_{52},U_{55},U_{63},U_{66}$} & \multicolumn{2}{l}{-1.7,99,-26,96} \\
    \midrule
    \multicolumn{3}{l}{Primary annihilation ($>1\%$)} & Fraction [\%] &
    \multicolumn{3}{l}{Primary annihilation ($>1\%$)} & Fraction [\%] \\
    \multicolumn{3}{l}{$\tilde{\chi}^0_1\,\tilde{\chi}_{1}^{-}\rightarrow d\bar{u}/s\bar{c}/b\bar{t}/\bar{\nu}_e e/\bar{\nu}_\mu \mu/\bar{\nu}_\tau \tau$} & 26.6 &
    \multicolumn{3}{l}{$\tilde{\chi}^0_1\,\tilde{\chi}_{1}^{-}\rightarrow d\bar{u}/s\bar{c}/b\bar{t}/\bar{\nu}_e e/\bar{\nu}_\mu \mu/\bar{\nu}_\tau \tau$} & 20.5 \\
    \multicolumn{3}{l}{$\tilde{\chi}^0_1\,\tilde{\chi}_{1}^{-}\rightarrow \gamma W_{1}^{-}/Z W_{1}^{-}$} & 2.5 &
    \multicolumn{3}{l}{$\tilde{\chi}^0_2\,\tilde{\chi}_{1}^{-}\rightarrow d\bar{u}/s\bar{c}/b\bar{t}/\bar{\nu}_e e/\bar{\nu}_\mu \mu/\bar{\nu}_\tau \tau$} & 17.6 \\
    \multicolumn{3}{l}{$\tilde{\chi}^0_2\,\tilde{\chi}_{1}^{-}\rightarrow d\bar{u}/s\bar{c}/b\bar{t}/\bar{\nu}_e e/\bar{\nu}_\mu \mu/\bar{\nu}_\tau \tau$} & 19.1 &
    \multicolumn{3}{l}{$\tilde{\chi}_{1}^{-}\,\tilde{\chi}_{1}^{+}\rightarrow u\bar{u}/c\bar{c}/t\bar{t}/d\bar{d}/s\bar{s}/W^{-}W^{+}$} & 8.6 \\
    \multicolumn{3}{l}{$\tilde{\chi}_{1}^{-}\,\tilde{\chi}_{1}^{+}\rightarrow u\bar{u}/c\bar{c}/t\bar{t}/d\bar{d}/s\bar{s}$} & 10.4 &
    \multicolumn{3}{l}{$\tilde{\chi}^0_1\,\tilde{\chi}^0_1\rightarrow W^{-}W^{+}/ZZ$} & 3.0 \\
    \multicolumn{3}{l}{$\tilde{\chi}_{1}^{-}\,\tilde{\chi}_{1}^{+}\rightarrow W^{-}W^{+}/e\bar{e}/\mu\bar{\mu}/\tau\bar{\tau}$} & 6.0 &
    \multicolumn{3}{l}{$\tilde{\chi}^0_1\,\tilde{\chi}^0_2\rightarrow u\bar{u}/c\bar{c}/d\bar{d}/s\bar{s}/b\bar{b}$} & 8.7 \\
    \multicolumn{3}{l}{$\tilde{\chi}^0_1\,\tilde{\chi}^0_1\rightarrow W^{-}W^{+}/ZZ$} & 5.1 &
    \multicolumn{3}{l}{$\tilde{l}_1^{-}\,\tilde{\chi}_{1}^{-}\rightarrow W^{-}\tau^{-}$} & 1.2 \\
    \multicolumn{3}{l}{$\tilde{\chi}^0_1\,\tilde{\chi}^0_2\rightarrow d\bar{d}/s\bar{s}/b\bar{b}/u\bar{u}/c\bar{c}/t\bar{t}$} & 12.1 &
    \multicolumn{3}{l}{$\tilde{l}_1^{+}\,\tilde{l}_1^{-}\rightarrow W^{-}W^{+}$} & 1.1 \\
    \multicolumn{3}{l}{$\tilde{\chi}^0_1\,\tilde{\chi}^0_2\rightarrow \bar{\nu}_e e/\bar{\nu}_\mu \mu/\bar{\nu}_\tau \tau$} & 3.1 &
    \multicolumn{3}{l}{$\tilde{\nu}_1\,\tilde{\nu}_1^{*}\rightarrow W^{-}W^{+}/ZZ$} & 2.4 \\
    & & & &
    \multicolumn{3}{l}{$\tilde{\nu}_2\,\tilde{\nu}_2^{*}\rightarrow W^{-}W^{+}/ZZ$} & 2.4 \\
    & & & &
    \multicolumn{3}{l}{$\tilde{\nu}_3\,\tilde{\nu}_3^{*}\rightarrow W^{-}W^{+}/ZZ$} & 2.4 \\
    \midrule
    \multicolumn{2}{l}{DM observable} & \multicolumn{2}{l|}{Value} &
    \multicolumn{2}{l}{DM observable} & \multicolumn{2}{l}{Value} \\
    \multicolumn{2}{l}{$\Omega h^2$} & \multicolumn{2}{l|}{0.118} &
    \multicolumn{2}{l}{$\Omega h^2$} & \multicolumn{2}{l}{0.116} \\
    \multicolumn{2}{l}{$\sigma^{\text{SI}}_{p,n}/(10^{-46}\text{cm}^2)$} & \multicolumn{2}{l|}{$1.82,1.84$} &
    \multicolumn{2}{l}{$\sigma^{\text{SI}}_{p,n}/(10^{-47}\text{cm}^2)$} & \multicolumn{2}{l}{$1.37,1.38$} \\
    \multicolumn{2}{l}{$\sigma^{\text{SD}}_{p,n}/(10^{-43}\text{cm}^2)$} & \multicolumn{2}{l|}{$1.05,8.02$} &
    \multicolumn{2}{l}{$\sigma^{\text{SD}}_{p,n}/(10^{-43}\text{cm}^2)$} & \multicolumn{2}{l}{$1.46,1.12$} \\
    \bottomrule
  \end{tabular}
  \caption{Same as Table~\ref{BMP12}, but for benchmark points P3 and P4. Point P3 is chosen from the allowed (colored) region in the third quadrant ($M_1,M_2<0$) of Figure~\ref{Figure4}, while P4 is taken from the fourth quadrant ($M_1>0,M_2<0$). Both points are consistent with all current constraints, including the latest LZ2024 direct detection limits.}
  \label{BMP34}
\end{table}

\section{\label{conclusion}Conclusion}

We performed a comprehensive analysis of higgsino-dominated dark matter within the MSSM framework, focusing on the implications of the latest LZ2024 direct-detection limits when slepton co-annihilation channels are taken into account. Including constraints from the Higgs sector, $B$-physics, and the DM relic density, we progressively applied the LZ2022 and LZ2024 SI and SD limits to illustrate their impact on the parameter space. Based on this analysis, we draw the following conclusions

\begin{itemize}
    \item \textbf{Relic density and mass constraints} The relic-density requirement by itself typically favors a higgsino DM mass around $1~\text{TeV}$, since the efficient annihilation of higgsino-like DM leads to an under-abundance at lower masses. Our study shows that, once slepton co-annihilation channels are included, the effective annihilation cross section is modified in such a way that the correct relic density can be achieved at significantly lower masses. Under the LZ2022 constraints, the allowed DM mass can be as low as $\sim 400~\text{GeV}$. After imposing the more stringent LZ2024 limits, this lower bound is raised to about $500~\text{GeV}$. While slepton co-annihilation remains subdominant compared to $\tilde{\chi}_1^\pm$ and $\tilde{\chi}_2^0$ co-annihilation over most of the sub-TeV region, its relative importance grows as the DM mass decreases towards $500~\text{GeV}$. In this mass window, even a subdominant slepton contribution can play a crucial role in satisfying the relic-density constraint. Thus, it is not the efficiency of slepton annihilations that lowers the required higgsino mass, but rather their inefficient coannihilation compared to higgsinos, which reduces the effective annihilation cross section.

    \item \textbf{Direct detection and interference effects} For DM–nucleon scattering, the absolute size of the SI and SD cross sections is suppressed by large gaugino masses ($M_1$ and $M_2$), but the relative sign between $M_1$ and $M_2$ has a pronounced impact through interference effects in the DM–Higgs and DM–$Z$ couplings. When $M_1$ and $M_2$ have the same sign, constructive interference enhances the SI cross section, leading to strong constraints from LZ2024. Conversely, when $M_1$ and $M_2$ have opposite signs, destructive interference suppresses the SI cross section, allowing a much larger portion of the parameter space to evade current bounds. In particular, we find that the entire first quadrant ($M_1, M_2 > 0$) is excluded by LZ2024, whereas viable points survive in the third quadrant ($M_1, M_2 < 0$) thanks to gaugino decoupling and proximity to spin-independent blind spots. This sharp contrast between same-sign and opposite-sign regions constitutes one of the key qualitative outcomes of our analysis.
\end{itemize}

Regarding collider constraints, the mass splittings between the higgsino-like LSP and the other higgsino states ($\tilde{\chi}_2^0$ and $\tilde{\chi}_1^\pm$) in our scenarios are typically very small (below about $1\%$), leading to soft decay products that are challenging to detect. The sleptons involved in co-annihilation are also nearly degenerate with the LSP, resulting in similarly soft final states. As discussed in Section~\ref{BMP}, a recast of the latest 13~TeV electroweakino searches with \texttt{CheckMATE2} yields $R \ll 1$ for all our benchmark points, so current LHC searches have limited sensitivity to such compressed spectra. Future collider experiments such as the HL-LHC or HE-LHC, as well as indirect searches  will be promissing to effectively probe the parameter space highlighted in this work.

\section*{Acknowledgement}
This work is supported by the National Natural Science Foundation of China (NNSFC) under grant No. 12575110 and Natural Science Foundation of Henan Province under Grant No. 232300421217. We sincerely thank Prof. Junjie Cao and D.r. Zhiyang Bao for helpful discussions during this project.

\appendix
\section{Dependence on Slepton Soft-Masses and Trilinear Couplings}

In the main text, we adopted a simplified slepton soft-mass ansatz with full flavor and left--right universality (characterized by a single parameter, $\Delta_L$) and set slepton trlinear coefficient $A_e(Q)=0$ to reduce the parameter space dimensionality and highlight the fundamental coannihilation mechanism with maximal clarity. To assess how our conclusions depend on this assumption, we perform an additional study in which we keep all other scan settings, tools, and constraints identical to the main text, and only modify the slepton soft-mass inputs. All soft inputs are specified at $Q=1~\mathrm{TeV}$.

\vspace{2mm}
We consider the following scenarios as robustness checks:
\begin{itemize}
  \item \textbf{Scenario A:} the first two generations are unified, while the third generation is allowed to differ; within each group we keep left--right universality. We parameterize the \emph{diagonal slepton soft masses} at $Q=1~\mathrm{TeV}$ as
  \begin{eqnarray}
    (m^2_{\tilde L})_{11}=(m^2_{\tilde L})_{22}=(m^2_{\tilde e})_{11}=(m^2_{\tilde e})_{22}
    &=|\mu|^2(1+\Delta_{L1})^2, \label{eq:A1}\\
    (m^2_{\tilde L})_{33}=(m^2_{\tilde e})_{33}
    &=|\mu|^2(1+\Delta_{L2})^2. \label{eq:A2}
  \end{eqnarray}

  \item \textbf{Scenario B:} the first two generations are the same as in Scenario A [Eq.~\eqref{eq:A1}], while the third-generation left- and right-handed soft masses are taken independent:
  \begin{eqnarray}
    (m^2_{\tilde L})_{33} &= |\mu|^2(1+\Delta_{L2})^2, \nonumber\\
    (m^2_{\tilde e})_{33} &= |\mu|^2(1+\Delta_{L3})^2. \label{eq:A3}
  \end{eqnarray}

  \item \textbf{Scenario C:} for four benchmark points we vary the diagonal charged-lepton trilinear inputs universally,
  \begin{eqnarray}
    A_{e,11}=A_{e,22}=A_{e,33}\equiv A_e = 1000,\ 2000,\ 3000~\mathrm{GeV},
\end{eqnarray}
  so that in particular $A_\tau=A_{e,33}=A_e$ to test the sensitivity of the spectrum to left--right mixing.
\end{itemize}

For Scenarios A and B we scan $\Delta_{L_i}\in[0,0.35]$ ($i=1,2,3$), matching the range used in the main text to characterize the slepton--LSP mass splitting. The introduction of additional parameters in these scenarios (two parameters in A, three in B, compared to one in the baseline) significantly expands the parameter space, challenging exhaustive sampling and allowing for a wider range of slepton mass hierarchies.

\vspace{2mm}
\noindent\textbf{Results for Scenarios A and B.}
For Scenarios A and B, we reproduce plots analogous to Fig.~1 and Fig.~2 of the main text.

\begin{figure}[htbp]
  \includegraphics[width=1.0\textwidth]{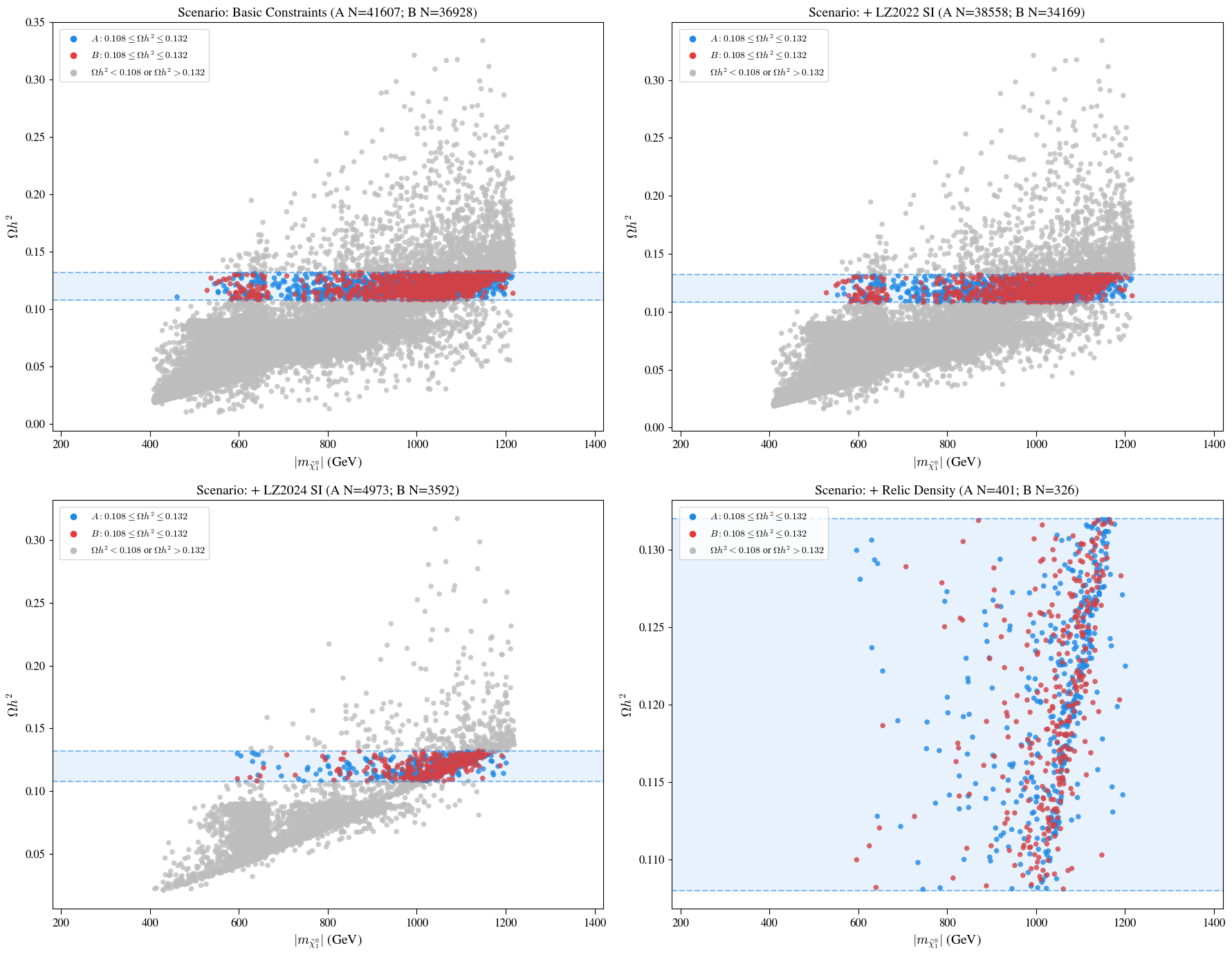}
  \caption{Relic density as a function of the DM mass for different slepton-mass assumptions. Red points: Scenario A. Blue points: Scenario B.\label{Figure1_Appendix}}
\end{figure}

As shown in Fig.~\ref{Figure1_Appendix}, Scenarios A and B still allow sub-TeV higgsino-like DM to satisfy the relic-density constraint, provided that the lightest slepton remains close in mass to the LSP. Thus, relaxing slepton universality in the ways considered here does not change the qualitative conclusion of our main text.

We note that the lowest viable masses in these extended scans appear at $\sim 600$~GeV, slightly higher than the baseline results $\sim 500$~GeV presented in Figure~\ref{Figure1}. We attribute this difference to scan coverage: Scenarios~A and B introduce additional degrees of freedom, significantly expanding the parameter space and increasing the computational challenge of exhaustive sampling. Consequently, we do not interpret this offset as a new physical threshold. Instead, both the baseline and extended scenarios support the same qualitative conclusion: higgsino-like DM with a mass well below 1~TeV is phenomenologically viable. The $\sim 500$~GeV limit from the baseline scan likely provides the more accurate estimate of the true kinematic boundary, benefiting from the higher sampling efficiency possible in the lower-dimensional parameter space.

\begin{figure}[htbp]
  \includegraphics[width=1.0\textwidth]{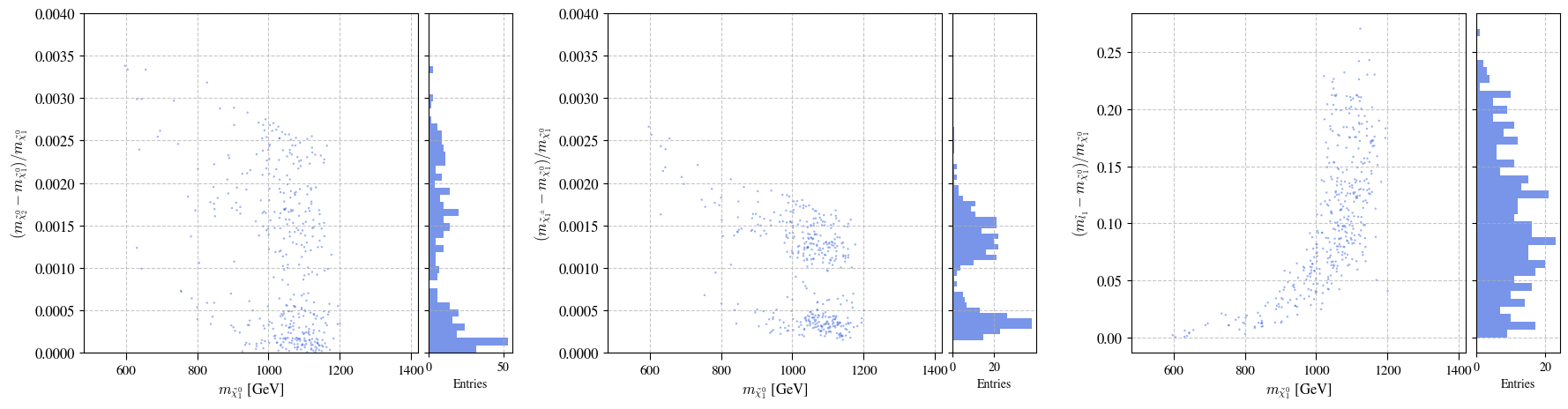}
  \includegraphics[width=1.0\textwidth]{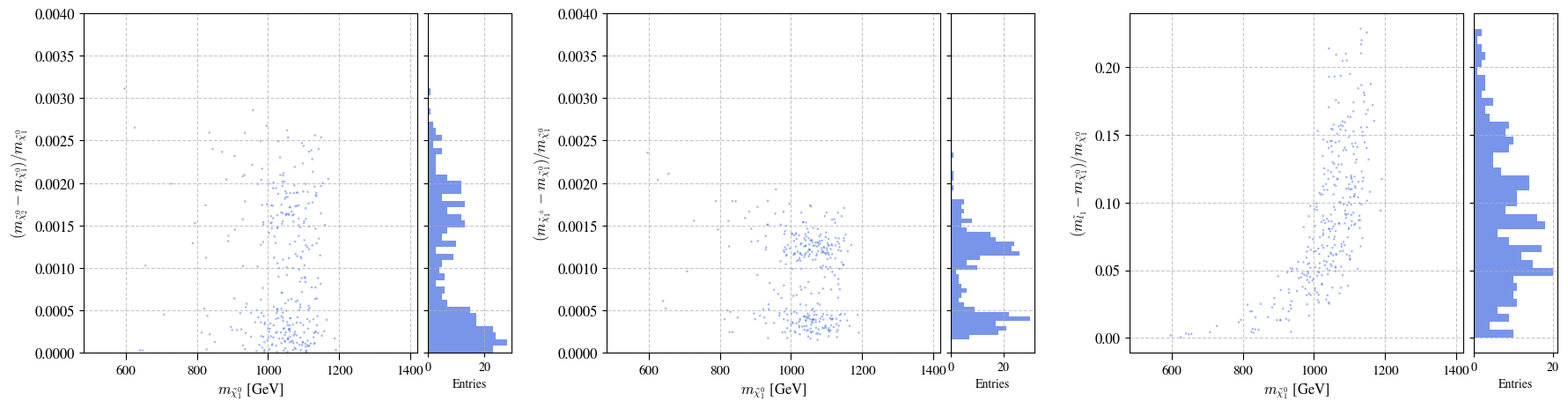}
  \caption{Mass splittings in Scenario A (upper panel) and Scenario B (lower panel). This figure should be compared with Fig.~2 in Sec.~3. The rightmost panels show the mass splitting between the lightest slepton-sector state $\tilde{l}_1$ and $\tilde{\chi}_1^0$, where $\tilde{l}_1$ is defined by $m_{\tilde{l}_1}\equiv \min\!\big(m_{\tilde{l}_1^\pm},\,m_{\tilde{\nu}_1}\big)$.\label{Figure2_Appendix}}
\end{figure}

Compared to the baseline scan result in Fig.~\ref{Figure2}, Scenarios A and B exhibit a broader and sparser distribution of viable points, particularly for the slepton--LSP mass splitting shown in the rightmost panels of Fig.~\ref{Figure2_Appendix}. This is an expected consequence of the increased parameter freedom in these scenarios, which allows for a wider range of slepton--DM mass hierarchies. While small $\Delta m$ values remain viable, the distribution of viable points is wider and extends to larger splittings. This observation, consistent with the expanded parameter space, demonstrates the robustness of the coannihilation mechanism across different slepton flavor and generation structures, even when the strictly degenerate limit is not uniquely favored by the enlarged parameter space.

We do not repeat the direct-detection plots for Scenarios A and B since, in our setup, the direct-detection rates are dominated by Higgs/$Z$ exchange and are only negligibly affected by changes in the slepton sector.

\vspace{2mm}
\noindent\textbf{Results for Scenario C.}
For Scenario C, we show the dependence of mass splittings on the trilinear input $A_e$ for our benchmark points P1--P4.

\begin{figure}[htbp]
  \includegraphics[width=1.0 \textwidth]{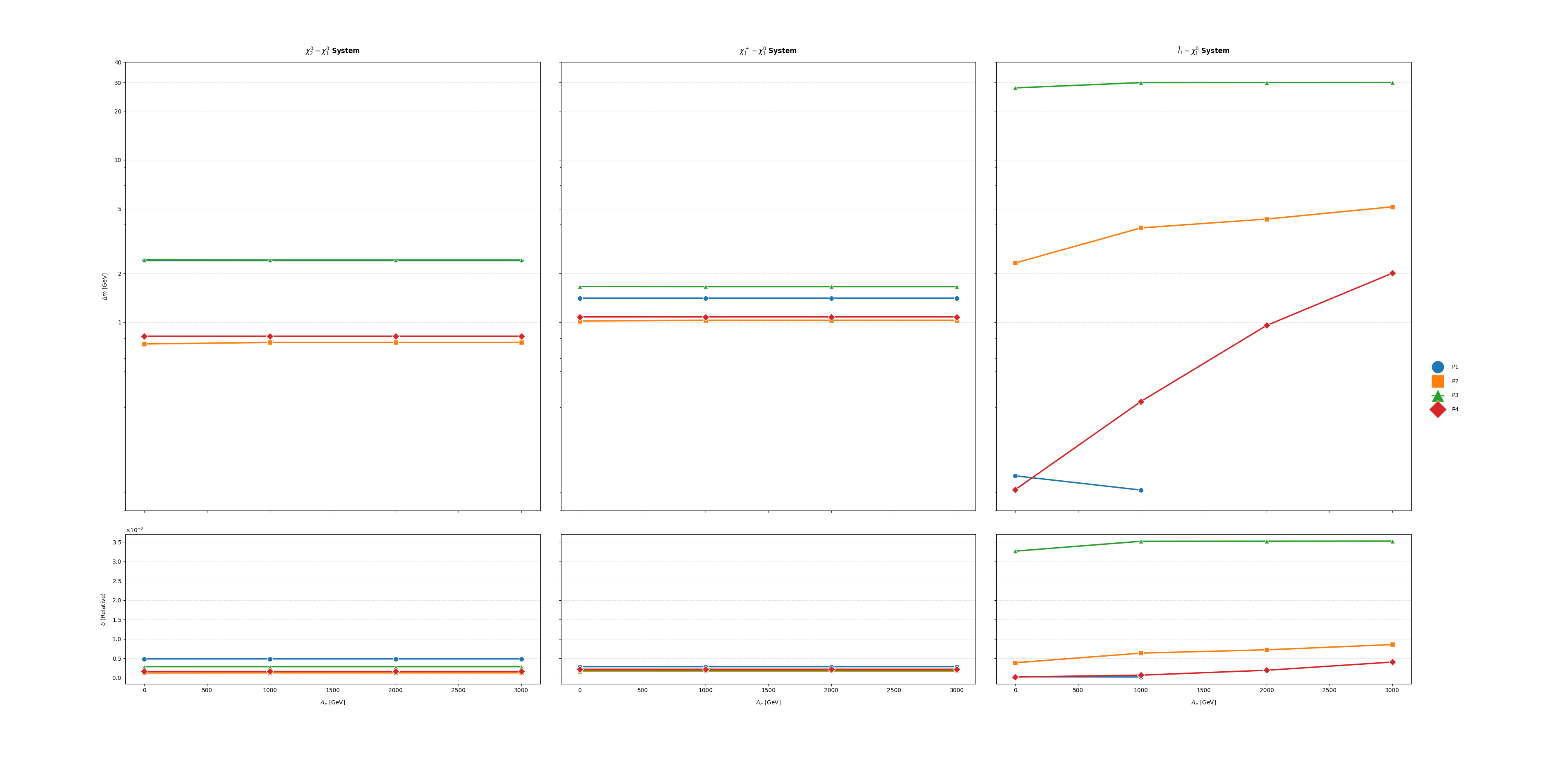}
  \caption{Mass splittings for Scenario C as a function of $A_e$. The upper panels show the absolute splittings for $m_{\tilde{\chi}_2^0}-m_{\tilde{\chi}_1^0}$ (left), $m_{\tilde{\chi}_1^\pm}-m_{\tilde{\chi}_1^0}$ (middle), and $m_{\tilde{l}_1}-m_{\tilde{\chi}_1^0}$ (right); note the logarithmic vertical axis. The label $\tilde{l}_1$ is a shorthand for the lightest slepton-sector coannihilating state, defined by $m_{\tilde{l}_1}\equiv \min\!\big(m_{\tilde{l}_1^\pm},\,m_{\tilde{\nu}_1}\big)$. The lower panels show the corresponding relative splittings $\delta=(m_i-m_{\tilde{\chi}_1^0})/m_{\tilde{\chi}_1^0}$ for three columns. \label{Figure3_Appendix}}
\end{figure}

As illustrated in Fig.~\ref{Figure3_Appendix}, the mass splittings exhibit mild sensitivities to variations in $A_e$. The upper panels, which display $\Delta m$ in GeV on a logarithmic y-axis, reveal that the $\tilde{\chi}_2^0$ -- $\tilde{\chi}_1^0$ and $\tilde{\chi}_1^\pm$--$\tilde{\chi}_1^0$ splittings are essentially insensitive to $A_e$. In contrast, the slepton–LSP splitting shows a more complex dependence.

\begin{itemize}
    \item For P1, the slepton--LSP splitting is initially small at low $A_e$. However, its trajectory becomes discontinuous around $A_e \approx 1\,~\text{TeV}$. This discontinuity occurs because, at larger $A_e$ values, the LSP undergoes a transition from a neutralino to a sneutrino. Consequently, our conclusions concerning Higgsino-dominated dark matter with slepton coannihilation are robust and primarily applicable up to $A_e \simeq 1\,\text{TeV}$. Beyond this point, the change in the LSP fundamentally alters the dark matter scenario, and the simple coannihilation picture we discuss no longer applies.

    \item For the remaining benchmark points, we observe diffferent trends in the slepton--LSP splitting. For P2, the mass splitting starts at approximately 2 GeV for $A_e=0$ and gradually increases to about 5 GeV by $A_e=3\,\text{TeV}$. Meanwhile, P3 exhibits a consistently larger splitting of around 30 GeV, which remains nearly constant across the entire $A_e$ range. Lastly, for P4, the splitting begins with a very small value, close to 0 GeV, and increases to roughly 2 GeV as $A_e$ reaches 3 TeV.
\end{itemize}

The lower panels of Fig.~\ref{Figure3_Appendix} consistently demonstrate that the relative mass splittings remain well within the $\mathcal{O}(10\%)$ range for all relevant $A_e$ values. This observation strongly suggests that coannihilation processes continue to play a significant role, which is consistent with our primary conclusions. The overall mild dependence on $A_e$ is anticipated because, while $A_e$ directly contributes to the stau left-right mixing term $\mathcal{M}^2_{\tilde\tau,LR}=m_\tau(A_e-\mu\tan\beta)$, in our specific parameter region, this term's magnitude is predominantly governed by the much larger $\mu\tan\beta$ contribution. This renders variations due to $A_e$ comparatively small.


\bibliographystyle{CitationStyle}
\bibliography{myrefs}
\end{document}